\documentclass[reprint,prd,twocolumn,superscriptaddress]{revtex4-2}
\pdfoutput=1
\usepackage{times,mathptmx,graphicx,hyperref}
\begin{document}
\title{Correlators of heavy-light quark currents in HQET:\\
Perturbative contribution up to 4 loops and beyond}
\author{Andrey G.~Grozin}
\email{A.G.Grozin@inp.nsk.su}
\affiliation{Budker Institute of Nuclear Physics, Lavrentiev St.~11, Novosibirsk 630090, Russia}
\affiliation{Bogoliubov Laboratory of Theoretical Physics, Joint Institute for Nuclear Research,
Dubna 141980, Russia}
\date{}
\begin{abstract}
The perturbative contribution to the correlator of two HQET heavy-light currents
expanded in light-quark masses up to quadratic terms is calculated up to 4 loops.
The leading large-$\beta_0$ limit is also considered,
so that terms with the highest degrees of $n_f$ are calculated to all orders in $\alpha_s$.
Borel images of coefficient functions in this limit contain renormalon poles.
Naive nonabelianization works surprisingly poorly for the coefficient functions considered here.
\end{abstract}
\maketitle

\section{Introduction}
\label{S:Intro}

QCD problems with a single heavy quark (having the pole mass $M$)
can be handled by Heavy Quark Effective Theory
(HQET, see, e.g., \cite{Neubert:1993mb,Manohar:2000dt,Grozin:2004yc}).
At the leading order in $1/M$ the heavy quark spin does not interact with the gluon field
(chromomagnetic interaction is $\sim1/M$),
so that the heavy quark spin can be switched off.
The HQET Lagrangian with the fields $h$ (spin $\frac{1}{2}$) and $\varphi$ (spin 0)
in the static-quark rest frame,
\begin{equation}
L = h^+ i D_0 h + \varphi^* i D_0 \varphi\,,
\label{L}
\end{equation}
has superflavor symmetry $SU(3)$~\cite{Georgi:1990ak}.
The coordinate-space propagator of the spinless heavy quark is
\begin{equation}
S_0(x) = \delta(\vec{x}^{\,}) S_0(x^0)\,,\quad
S_0(t) = - i \theta(t)\,,
\label{Scoord}
\end{equation}
i.e., the quark stays where it has been created;
it propagates only forward in time,
so that its line cannot form loops (it has no antiparticle).
The momentum-space propagator
\begin{equation}
S_0(p) = \frac{1}{p^0 + i0}
\label{Smom}
\end{equation}
depends only on $p^0$, but not on $\vec{p}$.

We consider heavy-light quark currents
\begin{equation}
j_{P0} = \frac{1 + P\gamma^0}{2} \varphi^*_0 q_0\,,\quad
P = \pm1\,,
\label{j}
\end{equation}
where $\varphi^*$ is the spinless heavy antiquark field,
$q$ is the field of a light quark,
$P$ is the current parity,
the index 0 means unrenormalized quantities.
The $P = +1$ current has the quantum numbers of the $\frac{1}{2}^+$ meson
which forms a superflavor symmetry triplet together with the $B$, $B^*$ mesons,
they have identical properties at the leading order in $1/M$;
the $P = -1$ has the quantum numbers of the $P$-wave $\frac{1}{2}^-$ meson
which forms another superflavor symmetry triplet together with the $P$-wave $0^+$, $1^+$ mesons.

The correlators are defined as
\begin{eqnarray}
&&\left\langle T j_{P0}(x) \bar{j}_{P0}(0) \right\rangle
= \delta(\vec{x}^{\,}) P \frac{1 + P\gamma^0}{2} \Pi_{P0}(x^0)\,,
\label{Pi0}\\
&&\Pi_{P0}(\omega) = \int_0^\infty dt\,\Pi_{P0}(t) e^{i \omega t}\,,
\nonumber\\
&&\Pi_{P0}(t) = \int_{-\infty}^{+\infty} \frac{d\omega}{2\pi} \Pi_{P0}(\omega) e^{-i \omega t}\,.
\nonumber
\end{eqnarray}
Correlators of the currents $\bar{h}_0 \Gamma q_0$ for any Dirac matrices $\Gamma$
can be trivially obtained from $\Pi_{P0}$.
Analytically continuing $\Pi_{P0}(t)$ from the $t>0$ half-axis to the $t = -i\tau$, $\tau>0$ half-axis
we obtain the Euclidean correlators $\Pi_{P0}(\tau)$.
The spectral density is defined as
\begin{equation}
\rho_{P0}(\omega) = \frac{1}{2 \pi} \left[\Pi_{P0}(\omega+i0) - \Pi_{P0}(\omega-i0)\right]\,.
\label{rho0}
\end{equation}

The operator product expansion (OPE) for the correlators is
\begin{eqnarray}
\Pi_{P0}(\tau) = \sum_{\mathcal{O}} C_{P\mathcal{O}0}(\tau) \langle\mathcal{O}_0\rangle\quad
\biggl(\tau \ll \frac{1}{\Lambda_{\overline{\text{MS}}}}\biggr)\,,
\label{OPE1}\\
\Pi_{P0}(\omega) = \sum_{\mathcal{O}} C_{P\mathcal{O}0}(\omega) \langle\mathcal{O}_0\rangle\quad
(- \omega \gg \Lambda_{\overline{\text{MS}}})\,.
\label{OPE2}
\end{eqnarray}
We include the light-quark masses to the operators $\mathcal{O}$.
Wilson coefficients of even-dimensional operators $\mathcal{O}$ don't depend on $P$;
those of odd-dimensional operators are proportional to $P$.
In what follows, we set $P = +1$;
results for $P = -1$ can be reconstructed trivially.
Here we consider the operators up to dimension $D=2$:
\begin{equation}
\mathcal{O}_0 = 1\,,\quad
m_0\,,\quad
m_0^2\,,\quad
\sum m_{i0}^2
\label{O}
\end{equation}
(times the unit operator),
where $m$ is the mass of the quark field $q$ in our current $j_P$~(\ref{j}),
and $m_i$ are the masses of all light flavors.
These contributions are perturbative:
we just expand integrands of bare perturbative diagrams for the correlator in $m_0$, $m_{i0}$.
At $D=3$ there is the light-quark condensate contribution
(its calculation needs diagrams with 2 external light-quark lines having vanishing momenta)
and cubic mass contributions.
Both kinds of contributions are IR divergent; these divergences cancel.
They mix under renormalization (up to 3 loops they have been calculated in~\cite{Chetyrkin:2021qvd}).

We use $\overline{\text{MS}}$ renormalization scheme, $\mu$ is the normalization scale:
\begin{eqnarray}
&&\frac{g_0^2}{(4\pi)^{d/2}} = Z_\alpha \frac{\alpha_s(\mu)}{4\pi} \mu^{2\varepsilon} e^{\gamma_E\varepsilon}\,,
\nonumber\\
&&\beta = \frac{1}{2} \frac{d\log Z_\alpha}{d\log\mu} = \sum_{l=1}^\infty \beta_{l-1} \biggl(\frac{\alpha_s}{4\pi}\biggr)^l\,,
\nonumber\\
&&j_{P0} = Z_j j_P(\mu)\,,\quad
m_0 = Z_m m(\mu)\,,
\nonumber\\
&&\gamma_a = \frac{d\log Z_a}{d\log\mu} = \sum_{l=1}^\infty \gamma_{a,l-1} \biggl(\frac{\alpha_s}{4\pi}\biggr)^l\,,\quad
a = j, m;
\label{gammabeta}
\end{eqnarray}
The correlator of the renormalized currents is expressed via the renormalized condensates by OPE
\begin{eqnarray}
\Pi_P(\tau,\mu) = \sum_{\mathcal{O}} C_{P\mathcal{O}}(\tau,\mu) \langle\mathcal{O}(\mu)\rangle\quad
\biggl(\tau \ll \frac{1}{\Lambda_{\overline{\text{MS}}}}\biggr)\,,
\label{OPE1r}\\
\Pi_P(\omega,\mu) = \sum_{\mathcal{O}} C_{P\mathcal{O}}(\omega,\mu) \langle\mathcal{O}(\mu)\rangle\quad
(- \omega \gg \Lambda_{\overline{\text{MS}}})\,.
\label{OPE2r}
\end{eqnarray}
The coefficient functions $C_{P\mathcal{O}}(\mu)$ contain hard momenta $>\mu$ (short distances $<1/\mu$);
the matrix elements $\langle\mathcal{O}(\mu)\rangle$ contain soft momenta $<\mu$ (long distances $>1/\mu$).

The anomalous dimension of $C_{P,m^n}(\mu)$ ($n \in [0,2]$) is
\begin{equation}
\gamma_n = 2 \gamma_j - n \gamma_m\,.
\label{gamman}
\end{equation}
The solution of the renormalization group (RG) equation for the renormalized coefficient function $C_{P,m^n}(\tau,\mu)$
can be written as
\begin{eqnarray}
&&C_{P,m^n}(\tau,\mu) = \hat{C}_{P,m^n}(\tau)
\biggl(\frac{\alpha_s(\mu)}{\alpha_s(\mu_0)}\biggr)^{\!\!\gamma_{n,0}/(2\beta_0)} K_n(\alpha_s(\mu))\,,
\label{RG0}\\
&&K_n(\alpha_s) = \exp \int_0^{\alpha_s} \frac{d\alpha_s}{\alpha_s}
\biggl(\frac{\gamma_n(\alpha_s)}{2\beta(\alpha_s)} - \frac{\gamma_{n,0}}{2\beta_0}\biggr)\,,
\label{K}
\end{eqnarray}
where $\mu_0$ is some constant, and $\hat{C}_{P,m^n}(\tau)$ does not depend on $\mu$;
for $C_{P,m^n}(\omega,\mu)$ the solution is similar.
The $\mu$ dependence of the renormalized correlator $\Pi_{P}(\tau,\mu)$
(as well as that of $\Pi_{P}(\omega,\mu)$, $\rho_{P}(\omega,\mu)$)
is given by the formulas similar to~(\ref{RG0}) with $K_0(\alpha_s(\mu))$.

The correlator $\Pi_P(\omega,\mu)$ of the renormalized currents $j_P(\mu)$ contains
ultraviolet (UV) divergences $1/\varepsilon^n$;
their coefficients are polynomial in $\omega$.
Therefore, the coefficients of the divergences in $\Pi_P(t,\mu)$ contain $\delta^{(n)}(t)$.
They disappear in the analytical continuation to $\Pi_P(\tau,\mu)$;
divergences are also absent in the renormalized spectral density $\rho_P(\omega,\mu)$.

The correlator of two heavy-light quark currents in HQET
has been calculated at 2~\cite{Broadhurst:1991fc,Bagan:1991sg,Neubert:1991sp}
and 3~\cite{Chetyrkin:2021qvd} loops.
Here we consider the perturbative contribution at 4 loops (Sect.~\ref{S:4L}).
This correlator can be used in QCD sum rules for $f_B$, $f_{B^*}$, $\bar{\Lambda}$,
as well as for similar quantities for the $P$-wave $0^+$, $1^+$ mesons;
these topics will be discussed in Sect.~\ref{S:Conc}.

In Sect.~\ref{S:Lb0} we consider the large-$\beta_0$ limit.
This limit is a theoretical laboratory for investigating the high-order behavior of perturbative series,
including the ambiguities of their sums.
Its properties are qualitatively similar to those of full QCD,
but all-order perturbative results can be explicitly derived and analyzed.
There are, however, no good reasons to expect that these results are quantitatively applicable in full QCD;
this limit is just an instructive model.

\section{4-loop calculation}
\label{S:4L}

\subsection{Euclidean-time correlator}

We use the same calculation framework as in~\cite{Grozin:2023dlk}.
Diagrams are generated by \textsf{qgraf}~\cite{Nogueira:1991ex};
Dirac traces and contractions are calculated by \textsf{form}~\cite{Kuipers:2012rf,Ruijl:2017dtg};
color factors are calculated by the \textsf{form} package \textsf{color}~\cite{vanRitbergen:1998pn}.
IBP reduction is done by \textsf{LiteRed}~\cite{Lee:2012cn,Lee:2013mka} version 2,
the master integrals expanded in $\varepsilon$ have been obtained in~\cite{Lee:2022art}.
The correlator is gauge invariant.
We did the calculation exactly in the covariant-gauge parameter $\xi$ up to 3 loops,
and up to $\xi^1$ at 4 loops.
Cancellation of gauge-dependent terms is a strong check of our calculation.

Expanding the solution~(\ref{RG0}) of the RG equation
we obtain the structure of the coefficients $C_{m^n}(\tau,\mu)$:~%
\footnote{Note a typo in~\cite{Chetyrkin:2021qvd}, formula~(34):
the formula is correct if $\gamma_n$ is defined as $\gamma_j-\frac{n}{2}\gamma_m$,
not as~(\ref{gamman}).}
\begin{eqnarray}
&&C_{m^n}(\tau) = \frac{N_c C_0^{(m^n)}}{\pi^2 \tau^{3-n}}
\exp\biggl\{
\frac{\alpha_s}{4\pi}
(- \gamma_{n,0} L_\tau + c_1^{(m^n)})
\nonumber\\
&&{} + \biggl(\frac{\alpha_s}{4\pi}\biggr)^{\!\!2}
\bigl[- \beta_0 \gamma_{n,0} L_\tau^2 + (- \gamma_{n,1} + 2 \beta_0 c_1^{(m^n)}) L_\tau + c_2^{(m^n)}\bigr]
\nonumber\\
&&{} + \biggl(\frac{\alpha_s}{4\pi}\biggr)^{\!\!3}
\biggl[- \frac{4}{3} \beta_0^2 \gamma_{n,0} L_\tau^3
  + (- 2 \beta_0 \gamma_{n,1} - \beta_1 \gamma_{n,0} + 4 \beta_0^2 c_1^{(m^n)}) L_\tau^2
\nonumber\\
&&\quad{} + (- \gamma_{n,2} + 4 \beta_0 c_2^{(m^n)} + 2 \beta_1 c_1^{(m^n)}) L_\tau
  + c_3^{(m^n)}\biggr]
+ \cdots
\biggr\}\,,
\label{RG}
\end{eqnarray}
where
\begin{equation}
L_\tau = \log\frac{\mu}{\mu_{\tau0}}\,,\quad
\mu_{\tau0} = \frac{2}{\tau} e^{-\gamma_E}\,.
\label{mut0}
\end{equation}
We need $\gamma_j$ up to 3 loops~\cite{Chetyrkin:2003vi}
(recently is has been calculated at 4 loops~\cite{Grozin:2023dlk});
$\gamma_m$ up to 3 loops and $\beta$ up to 2 loops are well known.
The results are
\begin{widetext}
\begin{eqnarray}
C_0^{(1)} &=& \frac{1}{2}\,,\quad
c_1^{(1)} = 4 C_F \biggl(\frac{\pi^2}{3} + 2\biggr)\,,
\nonumber\\
c_2^{(1)} &=& - C_F \biggl[
C_F \biggl(8 \zeta_3 + \frac{32}{45} \pi^4 - \frac{20}{3} \pi^2 + \frac{103}{8}\biggr)
+ C_A \biggl(104 \zeta_3 + \frac{8}{45} \pi^4 + \frac{5}{27} \pi^2 - \frac{6413}{72}\biggr)
- T_F n_f \biggl(32 \zeta_3 + \frac{16}{27} \pi^2 - \frac{589}{18}\biggr)
\biggr]\,,
\nonumber\\
c_3^{(1)} &=& C_F \biggl[
C_F^2 \biggl(\frac{460}{3} \zeta_5 - 48 \zeta_3^2 - \frac{448}{9} \pi^2 \zeta_3 + \frac{292}{3} \zeta_3
+ \frac{2024}{2835} \pi^6 - \frac{109}{135} \pi^4 - \frac{320}{9} \pi^2 + \frac{763}{12}\biggr)
\nonumber\\
&&{} + C_F C_A \biggl(\frac{2434}{3} \zeta_5 + 48 \zeta_3^2 - \frac{400}{9} \pi^2 \zeta_3 - \frac{9182}{27} \zeta_3
- \frac{242}{2835} \pi^6 - \frac{15091}{810} \pi^4 + \frac{31064}{243} \pi^2 - \frac{124721}{648}\biggr)
\nonumber\\
&&{} + \frac{C_A^2}{3} \biggl(2086 \zeta_5 - \frac{548}{3} \pi^2 \zeta_3 - \frac{19504}{9} \zeta_3
+ \frac{458}{945} \pi^6 + \frac{18323}{270} \pi^4 - \frac{11722}{81} \pi^2 + \frac{2785705}{648}\biggr)
\nonumber\\
&&{} - \frac{C_F T_F n_f}{3} \biggl(1120 \zeta_5 - \frac{448}{3} \pi^2 \zeta_3 - \frac{10360}{9} \zeta_3
- \frac{1972}{135} \pi^4 + \frac{12148}{81} \pi^2 + \frac{15313}{27}\biggr)
\nonumber\\
&&{} - \frac{C_A T_F n_f}{3} \biggl(560 \zeta_5 + \frac{32}{3} \pi^2 \zeta_3 - \frac{2096}{9} \zeta_3
+ \frac{6154}{135} \pi^4 - \frac{1208}{9} \pi^2 + \frac{235895}{81}\biggr)
\nonumber\\
&&{} + \frac{2}{9} (T_F n_f)^2 \biggl(176 \zeta_3 + \frac{512}{45} \pi^4 - \frac{1120}{27} \pi^2 + \frac{18793}{27}\biggr)
\biggr]\,,
\label{m0}\\
C_0^{(m)} &=& \frac{1}{4}\,,\quad
c_1^{(m)} = 4 C_F \biggl(\frac{\pi^2}{3} + 3\biggr)\,,
\nonumber\\
c_2^{(m)} &=& - C_F \biggl[
C_F \biggl(20 \zeta_3 + \frac{32}{45} \pi^4 - \frac{4}{3} \pi^2 + \frac{55}{4}\biggr)
+ C_A \biggl(116 \zeta_3 + \frac{8}{45} \pi^4 - \frac{31}{27} \pi^2 - \frac{4981}{36}\biggr)
- T_F n_f \biggl(32 \zeta_3 + \frac{16}{27} \pi^2 - \frac{401}{9}\biggr)
\biggr]\,,
\nonumber\\
c_3^{(m)} &=& C_F \biggl[
C_F^2 \biggl(\frac{880}{3} \zeta_5 - 48 \zeta_3^2 - \frac{448}{9} \pi^2 \zeta_3 + \frac{688}{3} \zeta_3
+ \frac{2024}{2835} \pi^6 - \frac{661}{135} \pi^4 + \frac{64}{9} \pi^2 + \frac{2249}{6}\biggr)
\nonumber\\
&&{} + C_F C_A \biggl(\frac{2044}{3} \zeta_5 + 48 \zeta_3^2 - \frac{520}{9} \pi^2 \zeta_3 + \frac{10294}{27} \zeta_3
- \frac{242}{2835} \pi^6 - \frac{10417}{810} \pi^4 + \frac{22028}{243} \pi^2 - \frac{172129}{324}\biggr)
\nonumber\\
&&{} + \frac{C_A^2}{3} \biggl(2236 \zeta_5 - \frac{560}{3} \pi^2 \zeta_3 - \frac{32788}{9} \zeta_3
+ \frac{458}{945} \pi^6 + \frac{16157}{270} \pi^4 - \frac{7114}{81} \pi^2 + \frac{2192149}{324}\biggr)
\nonumber\\
&&{} - \frac{2}{3} C_F T_F n_f \biggl(560 \zeta_5 - \frac{224}{3} \pi^2 \zeta_3 - \frac{3200}{9} \zeta_3
- \frac{872}{135} \pi^4 + \frac{6650}{81} \pi^2 + \frac{10613}{27}\biggr)
\nonumber\\
&&{} - \frac{2}{3} C_A T_F n_f \biggl(280 \zeta_5 + \frac{16}{3} \pi^2 \zeta_3 - \frac{1786}{9} \zeta_3
+ \frac{2921}{135} \pi^4 - \frac{548}{9} \pi^2 + \frac{159647}{81}\biggr)
\nonumber\\
&&{} + \frac{4}{9} (T_F n_f)^2 \biggl(112 \zeta_3 + \frac{256}{45} \pi^4 - \frac{560}{27} \pi^2 + \frac{10573}{27}\biggr)
\biggr]\,,
\label{m1}\\
C_0^{(m^2)} &=& - \frac{1}{8}\,,\quad
c_1^{(m^2)} = 6 C_F\,,
\nonumber\\
c_2^{(m^2)} &=& C_F \biggl[C_F \biggl(40 \zeta_3 - \frac{32}{45} \pi^4 - 12 \pi^2 + \frac{363}{8}\biggr)
- C_A \biggl(22 \zeta_3 + \frac{4}{15} \pi^4 + \pi^2 - \frac{2789}{24}\biggr)
- \frac{229}{6} T_F n_f
\biggr]\,,
\nonumber\\
c_3^{(m^2)} &=& C_F \biggl[
- C_F^2 \biggl(\frac{5540}{3} \zeta_5 + 36 \zeta_3^2 - \frac{3008}{9} \pi^2 \zeta_3 - \frac{4630}{3} \zeta_3
- \frac{7232}{2835} \pi^6 + \frac{1421}{27} \pi^4 + \frac{1652}{9} \pi^2 - \frac{499}{4}\biggr)
\nonumber\\
&&{} + C_F C_A \biggl(\frac{94}{3} \zeta_5 + 192 \zeta_3^2 + \frac{1756}{9} \pi^2 \zeta_3 + \frac{62242}{27} \zeta_3
- \frac{2323}{5670} \pi^6 - \frac{27283}{810} \pi^4 - \frac{19138}{243} \pi^2 - \frac{79739}{648}\biggr)
\nonumber\\
&&{} + C_A^2 \biggl(\frac{3176}{3} \zeta_5 - 93 \zeta_3^2 - \frac{398}{9} \pi^2 \zeta_3 - \frac{14368}{27} \zeta_3
- \frac{17}{3780} \pi^6 + \frac{1103}{810} \pi^4 - \frac{10466}{243} \pi^2 + \frac{398035}{216}\biggr)
\nonumber\\
&&{} + \frac{C_F T_F n_f}{3} \biggl(1040 \zeta_5 - 320 \pi^2 \zeta_3 - \frac{18224}{9} \zeta_3
+ \frac{4876}{135} \pi^4 + \frac{12656}{81} \pi^2 - \frac{26545}{27}\biggr)
\nonumber\\
&&{} - \frac{C_A T_F n_f}{3} \biggl(1000 \zeta_5 - \frac{160}{3} \pi^2 \zeta_3 - \frac{776}{9} \zeta_3
+ \frac{34}{27} \pi^4 - \frac{1120}{81} \pi^2 + \frac{83383}{27}\biggr)
+ 2 (T_F n_f)^2 \biggl(16 \zeta_3 + \frac{5329}{81}\biggr)
\biggr]\,.
\label{m2}
\end{eqnarray}
\end{widetext}
The coefficients $c_{1,2}^{(m^n)}$ agree with~\cite{Chetyrkin:2021qvd}, $c_3^{(m^n)}$ are new.
The highest weight at $L$ loops is $2(L-1)$, at least up to $L=4$.
Numerically, in QCD with $n_f=4$, setting $\mu=\mu_{\tau0}$, we have
\begin{eqnarray}
C_1(\tau) &=& \frac{3}{2 \pi^2 \tau^3} \biggl[1
+ 7.05316 \frac{\alpha_s}{\pi}
+ 10.1485 \biggl(\frac{\alpha_s}{\pi}\biggr)^{\!\!2}
\nonumber\\
&&{} + 125.943 \biggl(\frac{\alpha_s}{\pi}\biggr)^{\!\!3}
+ \cdots\biggr]\,,
\nonumber\\
C_m(\tau) &=& \frac{3}{4 \pi^2 \tau^2} \biggl[1
+ 8.38649 \frac{\alpha_s}{\pi}
+ 22.9273 \biggl(\frac{\alpha_s}{\pi}\biggr)^{\!\!2}
\nonumber\\
&&{} + 215.543 \biggl(\frac{\alpha_s}{\pi}\biggr)^{\!\!3}
+ \cdots\biggr]\,,
\nonumber\\
C_{m^2}(\tau) &=& - \frac{3}{8 \pi^2 \tau} \biggl[1
+ 2 \frac{\alpha_s}{\pi}
- 1.35353 \biggl(\frac{\alpha_s}{\pi}\biggr)^{\!\!2}
\nonumber\\
&&{} + 150.042 \biggl(\frac{\alpha_s}{\pi}\biggr)^{\!\!3}
+ \cdots\biggr]\,.
\label{Ctnum}
\end{eqnarray}
We see that the coefficients of the perturbative series grow very quickly.
This is a typical behavior for quantities which have a renormalon at $u=\frac{1}{2}$,
the closest possible position to the origin ---
it leads to the fastest possible growth of the coefficients (see Sect.~\ref{S:Lb0}).

The RG structure of $C_{\sum m_i^2}(\tau)$ is
\begin{eqnarray}
&&C_{\sum m_i^2}(\tau) = - \frac{2 N_c C_F T_F}{3 \pi^2 \tau} \biggl(\frac{\alpha_s}{4\pi}\biggr)^{\!\!2}
\Bigl\{c_0^{(\sum m_i^2)}
\nonumber\\
&&{} + \frac{\alpha_s}{4\pi} \Bigl[ (-\gamma_{2,0} + 4 \beta_0) c_0^{(\sum m_i^2)} L_\tau + c_1^{(\sum m_i^2)} \Bigr] + \cdots\Bigr\}\,.
\label{RG2}
\end{eqnarray}
We obtain
\begin{eqnarray}
&&c_0^{(\sum m_i^2)} = \pi^2 - 6\,,\quad
c_1^{(\sum m_i^2)} = - 6 C_F \biggl(18 \zeta_3 + \frac{\pi^4}{9} - \frac{7}{3} \pi^2 - 5\biggr)
\nonumber\\
&&{} + C_A \biggl(45 \zeta_5 - 3 \pi^2 \zeta_3 - 165 \zeta_3 + \frac{197}{3780} \pi^6 + \frac{2}{15} \pi^4 + \frac{113}{9} \pi^2
\nonumber\\
&&{} - \frac{175}{3}\biggr)
+ 48 T_F n_f \biggl(\zeta_3 - \frac{\pi^2}{27} - \frac{1}{9}\biggr)\,.
\label{sm2}
\end{eqnarray}
The coefficient $c_0^{(\sum m_i^2)}$ agrees with~\cite{Chetyrkin:2021qvd},
$c_1^{(\sum m_i^2)}$ is new.
The $C_A$ term in $c_1^{(\sum m_i^2)}$ contains diagrams with a light-quark triangle;
the $C_F$ and $T_F n_f$ terms contain only simpler diagrams with 2-point light-quark insertions
to gluon propagators.
Numerically at $\mu = \mu_{\tau0}$
\begin{equation}
C_{\sum m_i^2}(\tau) = - \frac{\pi^2 - 6}{12 \pi^2 \tau} \biggl(\frac{\alpha_s}{\pi}\biggr)^{\!\!2}
\biggl[1 - 9.14737 \frac{\alpha_s}{\pi} + \cdots\biggr]\,.
\label{C3tnum}
\end{equation}

The Wilson coefficients $C_{\mathcal{O}}(\omega)$ are in the attached file;
notations are explained in the comment at the top of this file.

\subsection{Spectral density}

Taking the discontinuity of~(\ref{OPE2}) at the cut we get the spectral density
\begin{equation}
\rho_0(\omega,\mu) = \sum_{\mathcal{O}} R_{\mathcal{O},0}(\omega,\mu) \langle \mathcal{O}_0 \rangle
\label{OPErho}
\end{equation}
(quark-hadron duality).
The RG structure of the coefficient functions $R_{m^n}(\omega)$ for $0 \le n \le 2$
is given by~(\ref{RG}) with the substitutions
\begin{eqnarray*}
&&C_0^{(m^n)} \to R_0^{(m^n)}\,,\quad
c_i^{(m^n)} \to r_i^{(m^n)}\,,\\
&&\frac{1}{\tau^{3-n}} \to \omega^{2-n}\,,\quad
L_\tau \to L_\omega = \log\frac{\mu}{2\omega}
\end{eqnarray*}
(note that $L_\omega$ in~\cite{Chetyrkin:2021qvd} is defined with the opposite sign).
The formulas for $r_i^{(m^n)}$ are shorter when expressed via $c_i^{(m^n)}$~(\ref{m0}--\ref{m2}):
\begin{widetext}
\begin{eqnarray}
&&R^{(1)}_0 = \frac{1}{4}\,,\quad
r^{(1)}_1 = c^{(1)}_1 + 9 C_F\,,\quad
r^{(1)}_2 = c^{(1)}_2 + C_F
\biggl[5 C_F (\pi^2 + 3) + C_A \biggl(9 \pi^2 + \frac{379}{2}\biggr) - 2 T_F n_f (2 \pi^2 + 35)\biggr]\,,
\nonumber\\
&&r^{(1)}_3 = c^{(1)}_3 + C_F \biggl[
C_F^2 \biggl(36 \zeta_3 - \frac{8}{3} \pi^4 + 13 \pi^2 + 99\biggr)
- C_F C_A \biggl(582 \zeta_3 + \frac{812}{27} \pi^4 - \frac{2500}{9} \pi^2 - \frac{1995}{2}\biggr)
\nonumber\\
&&\quad{}
- \frac{C_A^2}{9} \biggl(22330 \zeta_3 + \frac{1088}{9} \pi^4 - 539 \pi^2 - 50078\biggr)
+ C_F T_F n_f \biggl(336 \zeta_3 + \frac{1568}{135} \pi^4 - \frac{1052}{9} \pi^2 - 685\biggr)
\nonumber\\
&&\quad{}
+ \frac{4}{9} C_A T_F n_f \biggl(3466 \zeta_3 + \frac{964}{45} \pi^4 - 97 \pi^2 - 8942\biggr)
- \frac{8}{3} (T_F n_f)^2 \biggl(\frac{320}{3} \zeta_3 + \frac{16}{27} \pi^4 - 2 \pi^2 - \frac{751}{3}\biggr)
\biggr]\,,
\label{r0}\\
&&R^{(m)}_0 = \frac{1}{4}\,,\quad
r^{(m)}_1 = c^{(m)}_1 + 12 C_F\,,\quad
r^{(m)}_2 = c^{(m)}_2 - C_F \biggl[
10 C_F \biggl(\frac{2}{3} \pi^2 - 7\biggr)
- \frac{2}{3} C_A \biggl(\frac{5}{3} \pi^2 + 337\biggr)
+ \frac{8}{3} T_F n_f \biggl(\frac{\pi^2}{3} + 29\biggr)
\biggr]\,,
\nonumber\\
&&r^{(m)}_3 = c^{(m)}_3 - 2 C_F \biggl[
C_F^2 \biggl(252 \zeta_3 + \frac{40}{9} \pi^4 - \frac{70}{3} \pi^2 - 359\biggr)
+ C_F C_A \biggl(722 \zeta_3 + \frac{452}{27} \pi^4 + \frac{2753}{27} \pi^2 - \frac{15718}{9}\biggr)
\nonumber\\
&&\quad{}
+ \frac{C_A^2}{9} \biggl(9526 \zeta_3 + \frac{502}{9} \pi^4 + \frac{5507}{9} \pi^2 - 27238\biggr)
- 2 C_F T_F n_f \biggl(176 \zeta_3 + \frac{448}{135} \pi^4 + \frac{380}{27} \pi^2 - \frac{3637}{9}\biggr)
\nonumber\\
&&\quad{}
- \frac{4}{9} C_A T_F n_f \biggl(1354 \zeta_3 + \frac{458}{45} \pi^4 + \frac{1007}{9} \pi^2 - 4553\biggr)
+ \frac{16}{3} (T_F n_f)^2 \biggl(\frac{64}{3} \zeta_3 + \frac{4}{27} \pi^4 + \frac{47}{27} \pi^2 - 59\biggr)
\biggr]\,,
\label{r1}\\
&&R^{(m^2)}_0 = - \frac{1}{8}\,,\quad
r^{(m^2)}_1 = c^{(m^2)}_1\,,\quad
r^{(m^2)}_2 = c^{(m^2)}_2 - \pi^2 C_F (27 C_F + 11 C_A - 4 T_F n_f)\,,
\nonumber\\
&&r^{(m^2)}_3 = c^{(m^2)}_3 - C_F \biggl[
C_F^2 (1944 \zeta_3 + 16 \pi^4 + 3 \pi^2)
+ 2 C_F C_A \biggl(1188 \zeta_3 + \frac{34}{27} \pi^4 + \frac{1693}{9} \pi^2\biggr)
+ \frac{C_A^2}{3} \biggl(1936 \zeta_3 - \frac{44}{9} \pi^4 + \frac{1681}{3} \pi^2\biggr)
\nonumber\\
&&\quad{}
- 4 C_F T_F n_f \biggl(216 \zeta_3 + \frac{16}{27} \pi^4 + \frac{271}{9} \pi^2\biggr)
- \frac{4}{3} C_A T_F n_f \biggl(352 \zeta_3 - \frac{4}{9} \pi^4 + \frac{269}{3} \pi^2\biggr)
+ 16 (T_F n_f)^2 \biggl(\frac{16}{3} \zeta_3 + \pi^2\biggr)
\biggr]\,.
\label{r2}
\end{eqnarray}
\end{widetext}
The coefficients $r_{1,2}^{(m^n)}$ agree with~\cite{Chetyrkin:2021qvd},
$r_3^{(m^n)}$ are new.
Numerically, in QCD with $n_f=4$, setting $\mu=\mu_{\omega0}$,
\begin{equation}
\mu_{\omega0} = 2\omega\,,
\label{muw0}
\end{equation}
we have
\begin{eqnarray}
R_1(\omega) &=& \frac{3 \omega^2}{4 \pi^2} \biggl[1
+ 10.0532 \frac{\alpha_s}{\pi}
+ 94.2929 \biggl(\frac{\alpha_s}{\pi}\biggr)^{\!\!2}
\nonumber\\
&&{} + 939.006 \biggl(\frac{\alpha_s}{\pi}\biggr)^{\!\!3}
+ \cdots\biggr]\,,
\nonumber\\
R_m(\omega) &=& \frac{3 \omega}{4 \pi^2} \biggl[1
+ 12.3865 \frac{\alpha_s}{\pi}
+ 109.497 \biggl(\frac{\alpha_s}{\pi}\biggr)^{\!\!2}
\nonumber\\
&&{} + 787.826 \biggl(\frac{\alpha_s}{\pi}\biggr)^{\!\!3}
+ \cdots\biggr]\,,
\nonumber\\
R_{m^2}(\omega) &=& - \frac{3}{8 \pi^2} \biggl[1
+ 2 \frac{\alpha_s}{\pi}
- 51.5240 \biggl(\frac{\alpha_s}{\pi}\biggr)^{\!\!2}
\nonumber\\
&&{} - 799.215 \biggl(\frac{\alpha_s}{\pi}\biggr)^{\!\!3}
+ \cdots\biggr]\,.
\label{Rwnum}
\end{eqnarray}

For $\sum m_i^2$ we have
\begin{eqnarray}
&&R_{\sum m_i^2}(\omega) = \frac{N_c C_F T_F}{\pi^2} \biggl(\frac{\alpha_s}{4\pi}\biggr)^{\!\!2}
\Bigl\{c_0^{(\sum m_i^2)}
\nonumber\\
&&{} + \frac{\alpha_s}{4\pi} \Bigl[ (-\gamma_{2,0} + 4 \beta_0) c_0^{(\sum m_i^2)} L_\omega + c_1^{(\sum m_i^2)} \Bigr] + \cdots\Bigr\}\,,
\label{RG3}
\end{eqnarray}
where $c_i^{(\sum m_i^2)}$ are given by~(\ref{sm2}).
The first term agrees with~\cite{Chetyrkin:2021qvd},
the second one is new.
Numerically,
\begin{equation}
R_{\sum m_i^2}(\omega) = - \frac{\pi^2 - 6}{12 \pi^2} \biggl(\frac{\alpha_s}{\pi}\biggr)^{\!\!2}
\biggl[1 - 9.14737 \frac{\alpha_s}{\pi} + \cdots\biggr]\,.
\label{R3wnum}
\end{equation}

\section{Large $\beta_0$ limit}
\label{S:Lb0}

\subsection{Method of calculation}

In schemes based on dimensional regularization the wall separating hard and soft momenta is fuzzy.
Coefficient functions contain, in addition to the main hard contributions,
also soft contributions leading to IR renormalons and the corresponding ambiguities;
matrix elements contain, in addition to the main soft contributions,
also hard contributions leading to UV renormalons and the corresponding ambiguities.
They are artifacts of scale separation --- they are absent in the full correlator~(\ref{OPE1r}--\ref{OPE2r}).
If one changes the prescription for summing divergent perturbative series for a coefficient function
(thus changing the value of the sum),
one has to change the values of higher-dimensional vacuum condensates accordingly.
In HQET there is also an unusual UV renormalon in coefficient functions~\cite{Beneke:1994sw};
the corresponding ambiguity cancels the IR renormalon ambiguity of the residual mass term, as discussed later.
We use the large $\beta_0$ limit to investigate these questions.

In the large $\beta_0$ limit
(see, e.g., \cite{Beneke:1998ui}, Chapter~8 in~\cite{Grozin:2004yc})
the correlator of the heavy-light currents has been first investigated in~\cite{Beneke:1994sw}.
Coefficients in the perturbative series for the bare $C_{m^n,0}(\tau)$ ($n \in [0,2]$) are polynomials in $T_F n_f$:
\begin{equation}
A_{n0}(\tau) = \frac{C_{m^n,0}(\tau)}{C_{m^n,0}^{(1)}(\tau)}
= 1 + \sum_{l=1}^\infty \sum_{k=0}^{l-1} a'_{nlk} (T_F n_f)^k \biggl(\frac{g_0^2}{(4\pi)^{d/2}}\biggr)^{\!\!l}
\label{nf}
\end{equation}
(here $C_{m^n,0}^{(1)}(\tau)$ is the 1-loop contribution, it is convergent).
We can re-write these polynomials as polynomials in $\beta_0$:
\begin{equation}
A_{n0}(\tau) = 1 + \sum_{l=1}^\infty \sum_{k=0}^{l-1} a_{nlk} \beta_0^k \biggl(\frac{g_0^2}{(4\pi)^{d/2}}\biggr)^{\!\!l}\,.
\label{beta0}
\end{equation}
Note that $n_f$ can also appear, e.g., via $n_f d_F^{abcd}$, so that it is multiplied by a quartic Casimir,
then it cannot be expressed via $\beta_0$;
such terms first appears at the order $g_0^8$, i.e. $1/\beta_0^4$,
they are not relevant for us here.
Here we consider the large $\beta_0$ limit:
$\beta_0 \alpha_s \sim 1$, $1/\beta_0$ is our small parameter.
We consider the first order in $1/\beta_0$
(in some problems is appears possible to calculate $1/\beta_0^2$ corrections,
but only in problems containing some factor which simplifies the analysis considerably).

The unrenormalized quantity $A_{n0}(\tau)$ can be written in the form~\cite{Palanques-Mestre:1983ogz,Broadhurst:1992si}:
\begin{eqnarray}
&&A_{n0}(\tau) = 1 + \frac{C_F}{\beta_0} \sum_{l=1}^\infty \frac{F_n(\varepsilon,l\varepsilon)}{l}
\biggl[\frac{\beta_0 g_0^2}{(4\pi)^{d/2}} \biggl(\frac{\tau e^{\gamma_E}}{2}\biggr)^{\!\!2\varepsilon}
e^{-\gamma_E\varepsilon} \frac{D(\varepsilon)}{\varepsilon}\biggr]^l
\nonumber\\
&&{} + \mathcal{O}\biggl(\frac{1}{\beta_0^2}\biggr)\,.
\label{A0}
\end{eqnarray}
Here
\begin{equation}
D(\varepsilon) = 6 e^{\gamma_E\varepsilon} \frac{\Gamma(1+\varepsilon) \Gamma^2(2-\varepsilon)}{\Gamma(4-2\varepsilon)}
= 1 + \frac{5}{3} \varepsilon + \cdots
\label{De}
\end{equation}
Re-expressing this result via the renormalized coupling constant
\begin{equation}
b = \beta_0 \frac{\alpha_s(\mu)}{4\pi} \sim 1\,,
\label{b}
\end{equation}
we have
\begin{eqnarray}
&&A_{n0}(\tau) = 1 + \frac{C_F}{\beta_0} \sum_{l=1}^\infty \frac{F_n(\varepsilon,l\varepsilon)}{l}
\biggl[\frac{b}{\varepsilon+b} \biggl(\frac{\mu\tau e^{\gamma_E}}{2}\biggr)^{\!\!2\varepsilon} D(\varepsilon)\biggr]^l
\nonumber\\
&&{} + \mathcal{O}\biggl(\frac{1}{\beta_0^2}\biggr)\,.
\label{Ab}
\end{eqnarray}
It is convenient to set $\mu = \mu_\tau$:
\begin{equation}
\mu_\tau = \frac{2 e^{-\gamma_E}}{\tau} D(\varepsilon)^{-1/(2\varepsilon)} \to \frac{2}{\tau} e^{-\gamma_E-5/6}\,,
\label{mut}
\end{equation}
then
\begin{equation}
A_{n0}(\tau) = 1 + \frac{C_F}{\beta_0} \sum_{l=1}^\infty \frac{F_n(\varepsilon,l\varepsilon)}{l}
\biggl(\frac{b}{\varepsilon+b}\biggr)^{\!\!l} + \mathcal{O}\biggl(\frac{1}{\beta_0^2}\biggr)\,.
\label{Amut}
\end{equation}
The functions $F_n(\varepsilon,u)$ contain all information about the $1/\beta_0$ terms; they are regular at the origin:
\begin{equation}
F_n(\varepsilon,u) = \sum_{i=0}^\infty \sum_{j=0}^\infty F_{n,ij} \varepsilon^i u^j\,.
\label{Feu}
\end{equation}
Expanding also $(b/(\varepsilon+b))^l$ in $b$ we obtain quadruple sums for $A_{n0}(\tau)$.

\begin{figure}[h]
\begin{picture}(74,9.5)
\put(11,4.75){\makebox(0,0){\includegraphics{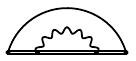}}}
\put(37,4.75){\makebox(0,0){\includegraphics{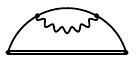}}}
\put(63,4.75){\makebox(0,0){\includegraphics{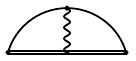}}}
\end{picture}
\caption{Two-loop diagrams}
\label{F:1}
\end{figure}

The functions $F_n(\varepsilon,u)$ can be found from the two-loop diagrams in Fig.~\ref{F:1}
where the denominator of the Landau-gauge gluon propagator is raised to the power $1+u-\varepsilon$.
Using integration by parts (a modification of an algorithm from~\cite{Grozin:2000jv}) we reduce them to $\Gamma$ functions
and a single non-trivial integral $I$:
\begin{widetext}
\begin{eqnarray}
&&F_0(\varepsilon,u) =
\frac{2 (3-2\varepsilon) \Gamma(2-2\varepsilon) e^{-2 \gamma_E u}}{3 (1-\varepsilon) \Gamma^2(1-\varepsilon) \Gamma(1+\varepsilon)} \biggl[
\frac{\bigl(3-2\varepsilon - (5+\varepsilon-2\varepsilon^2) u + (1+2\varepsilon) u^2\bigr) \Gamma(1-u)}%
{(1-2u) (1-\varepsilon-u) (2-\varepsilon-u) \Gamma(1-\varepsilon+u)}
\nonumber\\
&&\quad{} + \frac{2 u (\varepsilon+u) (1-2\varepsilon-2u)}{\Gamma(1-\varepsilon) \Gamma(1+2\varepsilon+2u)}
I(1+u-\varepsilon,\varepsilon)\biggr]\,,
\label{F0}\\
&&F_1(\varepsilon,u) =
\frac{2 (3-2\varepsilon) \Gamma(2-2\varepsilon) e^{-2 \gamma_E u}}{3 (1-\varepsilon) \Gamma^2(1-\varepsilon) \Gamma(1+\varepsilon)} \biggl[
\frac{\bigl(2 (3-2\varepsilon) - (14-5\varepsilon-2\varepsilon^2) u + (7-2\varepsilon) u^2\bigr) \Gamma(1-u)}%
{(1-2u) (1-\varepsilon-u) (2-\varepsilon-u) \Gamma(1-\varepsilon+u)}
\nonumber\\
&&\quad{} + \frac{2 u (\varepsilon+u) (1-2\varepsilon-2u)}{\Gamma(1-\varepsilon) \Gamma(1+2\varepsilon+2u)}
I(1+u-\varepsilon,\varepsilon)\biggr]\,,
\label{F1}\\
&&F_2(\varepsilon,u) =
\frac{2 (3-2\varepsilon) \Gamma(2-2\varepsilon) e^{-2 \gamma_E u}}{3 (1-\varepsilon) \Gamma^2(1-\varepsilon) \Gamma(1+\varepsilon)} \biggl[
\frac{\bigl(3 (3-2\varepsilon) - (24-11\varepsilon-2\varepsilon^2) u + (18-11\varepsilon+2\varepsilon^2) u^2 - (3-2\varepsilon) u^3\bigr)
\Gamma(1-u)}%
{(1+u) (1-2u) (1-\varepsilon-u) (2-\varepsilon-u) \Gamma(1-\varepsilon+u)}
\nonumber\\
&&\quad{} - \frac{4 \varepsilon u (\varepsilon+u)}{\Gamma(1-\varepsilon) \Gamma(1+2\varepsilon+2u)}
I(1+u-\varepsilon,\varepsilon)\biggr]\,,
\label{F2}
\end{eqnarray}
where~\cite{Beneke:1994sw}
\begin{eqnarray}
&&I(x,\varepsilon) = \int \frac{d^d k_1\,d^d k_2}{(i\pi^{d/2})^2} \frac{1}{(1-2k_1 v) (1-2k_2 v) (-k_1^2) (-k_2^2) \bigl[-(k_1-k_2)^2\bigr]^x}
\nonumber\\
&&{}=\frac{\Gamma\bigl(\frac{d}{2}-1\bigr) \Gamma\bigl(\frac{d}{2}-x-1\bigr)}{\Gamma(d-2)}
\biggl[2 \frac{\Gamma(2x-d+3) \Gamma(2(x-d+3))}{(x-d+3) \Gamma(3x-2d+6)}
{}_3F_2\biggl(\biggl.\begin{array}{c}x-d+3,x-d+3,2(x-d+3)\\x-d+4,3x-2d+6\end{array}\biggr|1\biggr)
\nonumber\\
&&\qquad{} - \Gamma(d-x-2) \Gamma^2(x-d+3)
\biggr]\,.
\label{BB}
\end{eqnarray}
\end{widetext}

Let's select $\varepsilon^{-1}$ terms from these sums for $A_n(\tau,\mu)$.
All coefficients $F_{n,ij}$ cancel except $F_{n,i0}$~\cite{Palanques-Mestre:1983ogz};
these terms, i.e. the functions $F_n(\varepsilon,0)$, give us the anomalous dimensions $\gamma_n$:
\begin{equation}
\gamma_n(b) = - 2 C_F \frac{b}{\beta_0} F_n(-b,0) + \mathcal{O}\biggl(\frac{1}{\beta_0^2}\biggr)\,.
\label{gamma}
\end{equation}
We have reproduced the result~\cite{Broadhurst:1994se}
\begin{eqnarray}
&&\gamma_j(b) = - \frac{1}{2} \gamma_m(b) + \mathcal{O}\biggl(\frac{1}{\beta_0^2}\biggr) =
\nonumber\\
&& - C_F \frac{b}{\beta_0} \frac{1+\frac{2}{3}b}{B(2+b,2+b) \Gamma(3+b) \Gamma(1-b)} + \mathcal{O}\biggl(\frac{1}{\beta_0^2}\biggr)\,,
\label{gammaj}
\end{eqnarray}
so that $\gamma_n = 2 (n+1) \gamma_j + \mathcal{O}(1/\beta_0^2)$.
In particular,
\begin{equation}
\gamma_{n,0} = - 2 C_F F_n(0,0)\,,\quad
F_n(0,0) = 3 (n+1)\,.
\label{gamma0}
\end{equation}
Terms of $A_{n0}(\tau)$ with $1/\varepsilon^{-n}$, $n>1$ contain no new information ---
they are unambiguously fixed by the $\varepsilon^{-1}$ terms
and the requirement that anomalous dimensions are finite at $\varepsilon\to0$;
hence they are also determined by $F_{n,i0}$.

The renormalized quantity $A_n(\tau,\mu)$ ($n \in [0,2]$) can be written in the form~(\ref{RG0}):
\begin{equation}
A_n(\tau,\mu) = \hat{A}_n(\tau) \biggl(\frac{\alpha_s(\mu)}{\alpha_s(\mu_\tau)}\biggr)^{\!\!\gamma_{n,0}/(2\beta_0)} K_n(\alpha_s(\mu))\,,
\label{RG1}
\end{equation}
where $\hat{A}_n(\tau)$ is an RG invariant ($\mu$ independent).
Let's select $\varepsilon^0$ terms from the quadruple sums for $A_{n0}(\tau)$.
All coefficients $F_{n,ij}$ cancel except $F_{n,i0}$ and $F_{n,0j}$~\cite{Broadhurst:1992si}.
The former ones (i.e. $F_n(-b,0)$) produce $K_n(b)$,
and the later ones (i.e. $F_n(0,u)$) produce $\hat{A}_n(\tau)$:
\begin{eqnarray}
&&\hat{A}_n(\tau) = 1 + \frac{C_F}{\beta_0} \int_0^\infty du\,e^{-u/b} S_n(u) + \mathcal{O}\biggl(\frac{1}{\beta_0^2}\biggr)\,,
\label{hatA}\\
&&S_n(u) = \frac{F_n(0,u)-F_n(0,0)}{u}
\label{Sn}
\end{eqnarray}
(here $b = b(\mu_\tau)$).
We have
\begin{eqnarray}
&&S_0(u) = \frac{1}{u} \biggl[\frac{2 (3-5u+u^2) \Gamma(1-u) e^{-2 \gamma_E u}}{(1-u) (2-u) (1-2u) \Gamma(1+u)} - 3\biggr]
\nonumber\\
&&\quad{} + \frac{4 u (1-2u) e^{-2 \gamma_E u}}{\Gamma(1+2u)} I(1+u,0)\,,
\label{S0}\\
&&S_1(u) = \frac{2}{u} \biggl[\frac{(6-14u+7u^2) \Gamma(1-u) e^{-2 \gamma_E u}}{(1-u) (2-u) (1-2u) \Gamma(1+u)} - 3\biggr]
\nonumber\\
&&\quad{} + \frac{4 u (1-2u) e^{-2 \gamma_E u}}{\Gamma(1+2u)} I(1+u,0)\,,
\label{S1}\\
&&S_2(u) = \frac{3}{u} \biggl[\frac{2 (3-5u+u^2) \Gamma(1-u) e^{-2 \gamma_E u}}{(1+u) (2-u) (1-2u) \Gamma(1+u)} - 3\biggr]\,.
\label{S2}
\end{eqnarray}
Here~(\ref{BB})
\begin{eqnarray}
&&I(1+u,0) = \frac{\Gamma(1-u)}{u^3}
\nonumber\\
&&\quad{}\times
\biggl[\Gamma(1-u) \Gamma^2(1+u) - \frac{\Gamma^2(1+2u)}{\Gamma(1+3u)} F(u)\biggr]\,,
\label{Fu1}\\
&&F(u) = {}_3F_2\biggl(\biggl.\begin{array}{c}u,u,2u\\1+u,1+3u\end{array}\biggr|1\biggr)
= 1
+ 2 \zeta_3 u^3
- \frac{4}{45} \pi^4 u^4
\nonumber\\
&&{} + 2 (2 \zeta_5 + \pi^2 \zeta_3) u^5
- 2 \biggl(11 \zeta_3^2 + \frac{79}{2835} \pi^6\biggr) u^6
\nonumber\\
&&{} - \biggl(\frac{167}{2} \zeta_7 - \frac{40}{3} \pi^2 \zeta_5 - \frac{79}{45} \pi^4 \zeta_3\biggr) u^7
\nonumber\\
&&{} + \biggl(\frac{44}{5} \zeta_{5,3} - 268 \zeta_3 \zeta_5 - 10 \pi^2 \zeta_3^2 - \frac{13387}{425250} \pi^8\biggr) u^8
\nonumber\\
&&{} + \cdots
\label{HypExp}\\
&&F\biggl(\frac{1}{2}\biggr) = \frac{3}{16} (\pi^2-4)\,,\quad
F(1) = \frac{3}{2}\,,
\nonumber\\
&&F(2) = \frac{25}{6}\,,\quad
F(3) = \frac{777}{50}\,.
\label{Fu2}
\end{eqnarray}
The expansion has been obtained using the \textsf{Mathematica} package \textsf{HypExp}~\cite{Huber:2005yg,Huber:2007dx}
(which uses \textsf{HPL}~\cite{Maitre:2005uu,Maitre:2007kp}).

Substituting the expansion
\begin{equation}
S_n(u) = \sum_{k=0}^\infty s_{nk} u^k
\label{Sex}
\end{equation}
into~(\ref{hatA}) we have
\begin{eqnarray}
&&\hat{A}_n(\tau) = 1 + \frac{C_F}{\beta_0} \sum_{l=1}^\infty \hat{c}_{nl} b^l
+ \mathcal{O}\biggl(\frac{1}{\beta_0^2}\biggr)\,,
\nonumber\\
&&\hat{c}_{nl} = (l-1)!\,s_{n,l-1}\,.
\label{Cex}
\end{eqnarray}

\subsection{Results for the euclidean-time correlator}

Up to 8 loops we obtain
\begin{widetext}
\begin{eqnarray}
&&\hat{A}_0(\tau) = 1 + C_F \frac{b}{\beta_0} \biggl[
\frac{4}{3} \pi^2 + \frac{11}{2}
- \biggl(24 \zeta_3 + \frac{8}{3} \pi^2 - \frac{39}{4}\biggr) b
+ \biggl(100 \zeta_3 +\frac{64}{45} \pi^4 + \frac{143}{4}\biggr) b^2
\nonumber\\
&&{} - \biggl(960 \zeta_5 - \frac{64}{3} \pi^2 \zeta_3 - 22 \zeta_3 + \frac{128}{15} \pi^4 - \frac{1629}{8}\biggr) b^3
+ \biggl(\frac{38544}{5} \zeta_5 - 960 \zeta_3^2 - \frac{512}{3} \pi^2 \zeta_3 + 156 \zeta_3
  + \frac{7216}{945} \pi^6 + \frac{6333}{4}\biggr) b^4
\nonumber\\
&&{} - \biggl(92400 \zeta_7 - 1024 \pi^2 \zeta_5 - 264 \zeta_5 - 9680 \zeta_3^2 - \frac{2048}{9} \pi^4 \zeta_3 - 1430 \zeta_3
  + \frac{14432}{189} \pi^6 - \frac{124785}{8}\biggr) b^5
\nonumber\\
&&{} - \biggl(25344 \zeta_{5,3} - \frac{7765920}{7} \zeta_7 + 314112 \zeta_3 \zeta_5 + 12288 \pi^2 \zeta_5 - 2808 \zeta_5
  - \frac{10240}{3} \pi^2 \zeta_3^2 - 880 \zeta_3^2 + \frac{8192}{3} \pi^4 \zeta_3 - 16290 \zeta_3
\nonumber\\
&&\qquad{} - \frac{168608}{1575} \pi^8 - \frac{1486035}{8}\biggr) b^6
+ \cdots\biggr]\,,
\label{hat0}\\
&&\hat{A}_1(\tau) = 1 + C_F \frac{b}{\beta_0} \biggl[
\frac{4}{3} \pi^2 + 7
- \biggl(24 \zeta_3 + \frac{8}{3} \pi^2 - \frac{21}{2}\biggr) b
+ \biggl(104 \zeta_3 + \frac{64}{45} \pi^4 + \frac{73}{2}\biggr) b^2
\nonumber\\
&&{} - \biggl(960 \zeta_5 - \frac{64}{3} \pi^2 \zeta_3 - 28 \zeta_3 + \frac{128}{15} \pi^4 - \frac{819}{4}\biggr) b^3
+ \biggl(\frac{38688}{5} \zeta_5 - 960 \zeta_3^2 - \frac{512}{3} \pi^2 \zeta_3 + 168 \zeta_3
  + \frac{7216}{945} \pi^6 + \frac{3171}{2}\biggr) b^4
\nonumber\\
&&{} - \biggl(92400 \zeta_7 - 1024 \pi^2 \zeta_5 - 336 \zeta_5 - 9760 \zeta_3^2 - \frac{2048}{9} \pi^4 \zeta_3 - 1460 \zeta_3
+ \frac{14432}{189} \pi^6 - \frac{62415}{4}\biggr) b^5
\nonumber\\
&&{} - \biggl(25344 \zeta_{5,3} - \frac{7770240}{7} \zeta_7 + 314112 \zeta_3 \zeta_5 + 12288 \pi^2 \zeta_5 - 3024 \zeta_5
- \frac{10240}{3} \pi^2 \zeta_3^2 - 1120 \zeta_3^2 + \frac{8192}{3} \pi^4 \zeta_3 - 16380 \zeta_3
\nonumber\\
&&\qquad{} - \frac{168608}{1575} \pi^8 - \frac{743085}{4}\biggr) b^6
+ \cdots\biggr]\,,
\label{hat1}\\
&&\hat{A}_2(\tau) = 1 + C_F \frac{b}{\beta_0} \biggl[
- \frac{3}{2}
+ \frac{57}{4} b
+ 3 \biggl(4 \zeta_3 + \frac{27}{4}\biggr) b^2
- 3 \biggl(2 \zeta_3-\frac{609}{8}\biggr) b^3
+ 3 \biggl(\frac{144}{5} \zeta_5 + 76 \zeta_3 + \frac{1857}{4}\biggr) b^4
\nonumber\\
&&{} - 3 \biggl(24 \zeta_5 - 80 \zeta_3^2 - 270 \zeta_3 - \frac{42885}{8}\biggr) b^5
+ 3 \biggl(\frac{4320}{7} \zeta_7 + 1368 \zeta_5 - 80 \zeta_3^2 + 6090 \zeta_3 + \frac{480015}{8}\biggr) b^6
+ \cdots\biggr]\,.
\label{hat2}
\end{eqnarray}
The expansion can be extended if desired using known algorithms for expanding the hypergeometric function~(\ref{HypExp}).

If we don't set $\mu=\mu_\tau$ (\ref{mut}), we should substitute
\begin{equation}
F_n(\varepsilon,u) \to F_n(\varepsilon,u,\mu) = F_n(\varepsilon,u) e^{2Lu}\,,\quad
L = \log\frac{\mu}{\mu_\tau}\,.
\label{Fmu}
\end{equation}
Setting $\mu=\mu_{\tau0}$~(\ref{mut0}) ($L = \frac{5}{6}$) we obtain
\begin{eqnarray}
&&A_0(\tau,\mu_{\tau0}) = 1 + C_F \frac{b_0}{\beta_0} \biggl[
4 \biggl(\frac{\pi^2}{3} + 2\biggr)
- \biggl(24 \zeta_3 + \frac{4}{9} \pi^2 - \frac{589}{24}\biggr) b_0
+ \biggl(22 \zeta_3 + \frac{64}{45} \pi^4 - \frac{140}{27} \pi^2 + \frac{18793}{216}\biggr) b_0^2
\nonumber\\
&&{} - \biggl(960 \zeta_5 - \frac{64}{3} \pi^2 \zeta_3 - \frac{1293}{4} \zeta_3
  + \frac{521}{360} \pi^4 + \frac{1300}{81} \pi^2 - \frac{31737}{64}\biggr) b_0^3
\nonumber\\
&&{} + \biggl(\frac{6562}{5} \zeta_5 - 960 \zeta_3^2 - \frac{256}{9} \pi^2 \zeta_3 + \frac{27427}{18} \zeta_3
  + \frac{7216}{945} \pi^6 - \frac{17929}{540} \pi^4 - \frac{9500}{243} \pi^2 + \frac{48811943}{12960}\biggr) b_0^4
\nonumber\\
&&{} - \biggl(92400 \zeta_7 - 1024 \pi^2 \zeta_5 - \frac{227039}{6} \zeta_5 - 1679 \zeta_3^2
  - \frac{2048}{9} \pi^4 \zeta_3 + \frac{22400}{27} \pi^2 \zeta_3 - \frac{1521929}{216} \zeta_3
\nonumber\\
&&\qquad{} + \frac{7219}{567} \pi^6 + \frac{665537}{3888} \pi^4 + \frac{62500}{729} \pi^2 - \frac{3411646627}{93312}\biggr) b_0^5
\nonumber\\
&&{} - \biggl(25344 \zeta_{5,3} - \frac{1297974}{7} \zeta_7 + 314112 \zeta_3 \zeta_5 + 2048 \pi^2 \zeta_5 - \frac{4279619}{18} \zeta_5
  - \frac{10240}{3} \pi^2 \zeta_3^2 - \frac{403755}{7} \zeta_3^2
\nonumber\\
&&\qquad{} + \frac{143351}{315} \pi^4 \zeta_3 + \frac{416000}{81} \pi^2 \zeta_3 - \frac{73858325}{1512} \zeta_3
\nonumber\\
&&\qquad{} - \frac{168608}{1575} \pi^8 + \frac{1767935}{3969} \pi^6 + \frac{85122241}{136080} \pi^4 + \frac{387500}{2187} \pi^2
  - \frac{93967251293}{217728}\biggr) b_0^6
+ \cdots\biggr] + \mathcal{O}\biggl(\frac{1}{\beta_0^2}\biggr)\,,
\label{lbC0}\\
&&A_1(\tau,\mu_{\tau0}) = 1 + C_F \frac{b_0}{\beta_0} \biggl[
4 \biggl(\frac{\pi^2}{3} + 3\biggr)
- \biggl(24 \zeta_3 + \frac{4}{9} \pi^2 - \frac{401}{12}\biggr) b_0
+ \biggl(28 \zeta_3 + \frac{64}{45} \pi^4 - \frac{140}{27} \pi^2 + \frac{10573}{108}\biggr) b_0^2
\nonumber\\
&&{} - \biggl(960 \zeta_5 - \frac{64}{3} \pi^2 \zeta_3 - \frac{701}{2} \zeta_3
  + \frac{53}{36} \pi^4 + \frac{1300}{81} \pi^2 - \frac{449939}{864}\biggr) b_0^3
\nonumber\\
&&{} + \biggl(\frac{6724}{5} \zeta_5 - 960 \zeta_3^2 - \frac{256}{9} \pi^2 \zeta_3 + \frac{14771}{9} \zeta_3
  + \frac{7216}{945} \pi^6 - \frac{8969}{270} \pi^4 - \frac{9500}{243} \pi^2 + \frac{2751007}{720}\biggr) b_0^4
\nonumber\\
&&{} - \biggl(92400 \zeta_7 - 1024 \pi^2 \zeta_5 - \frac{114463}{3} \zeta_5 - 1758 \zeta_3^2
  - \frac{2048}{9} \pi^4 \zeta_3 + \frac{22400}{27} \pi^2 \zeta_3 - \frac{271043}{36} \zeta_3
\nonumber\\
&&\qquad{}   + \frac{7222}{567} \pi^6 + \frac{332737}{1944} \pi^4 + \frac{62500}{729} \pi^2 - \frac{1712830147}{46656}\biggr) b_0^5
\nonumber\\
&&{} - \biggl(25344 \zeta_{5,3} - \frac{1302348}{7} \zeta_7 + 314112 \zeta_3 \zeta_5 + 2048 \pi^2 \zeta_5 - \frac{2159011}{9} \zeta_5
  - \frac{10240}{3} \pi^2 \zeta_3^2 - \frac{411030}{7} \zeta_3^2
\nonumber\\
&&\qquad{}   + \frac{143342}{315} \pi^4 \zeta_3 + \frac{416000}{81} \pi^2 \zeta_3 - \frac{12790375}{252} \zeta_3
\nonumber\\
&&\qquad{} - \frac{168608}{1575} \pi^8 + \frac{1767950}{3969} \pi^6 + \frac{42562241}{68040} \pi^4 + \frac{387500}{2187} \pi^2
- \frac{141101866679}{326592}\biggr) b_0^6
+ \cdots\biggr] + \mathcal{O}\biggl(\frac{1}{\beta_0^2}\biggr)\,,
\label{lbC1}\\
&&A_2(\tau,\mu_{\tau0}) = 1 + C_F \frac{b_0}{\beta_0} \biggl[
6 + \frac{229}{8} b_0 + \biggl(18 \zeta_3 + \frac{5329}{72}\biggr) b_0^2
+ \biggl(231 \zeta_3 - \frac{3}{10} \pi^4 + \frac{88673}{48}\biggr) \frac{b_0^3}{4}
\nonumber\\
&&{} + \biggl(\frac{486}{5} \zeta_5 + \frac{769}{2} \zeta_3 - \frac{\pi^4}{20} + \frac{15229703}{4320}\biggr) b_0^4
+ \biggl(\frac{1311}{2} \zeta_5 + 237 \zeta_3^2 + \frac{223369}{72} \zeta_3
  - \frac{\pi^6}{63} + \frac{7}{144} \pi^4 + \frac{1105270627}{31104}\biggr) b_0^5
\nonumber\\
&&{} + \biggl(\frac{13122}{7} \zeta_7 + \frac{13953}{2} \zeta_5 + \frac{15105}{7} \zeta_3^2
  + \frac{3}{35} \pi^4 \zeta_3 + \frac{6165575}{168} \zeta_3
  - \frac{5}{441} \pi^6 - \frac{83}{1680} \pi^4 + \frac{30732738973}{72576}\biggr) b_0^6
+ \cdots\biggr]
\nonumber\\
&&{} + \mathcal{O}\biggl(\frac{1}{\beta_0^2}\biggr)
\label{lbC2}
\end{eqnarray}
\end{widetext}
(here $b_0 = b(\mu_{\tau0})$).
Substituting $\beta_0 \to - \frac{4}{3} T_F n_f$,
we reproduce all $C_F (T_F n_f)^{l-1} \alpha_s(\mu_{\tau0})^l$ terms in (\ref{m0}--\ref{m2}),
and add further terms up to 8 loops (more such terms can be added if desired).
Leaving $\beta_0 = \frac{11}{3} C_A - \frac{4}{3} T_F n_f$ (naive nonabelianization~\cite{Broadhurst:1994se,Beneke:1994qe})
we get numerically with $n_f = 4$
\begin{eqnarray}
&&C_1(\tau,\mu_{\tau0}) = \frac{3}{2 \pi^2 \tau^3} \biggl[1
+ 7.05316 \frac{\alpha_s}{\pi}
- 6.03763 \biggl(\frac{\alpha_s}{\pi}\biggr)^{\!\!2}
\nonumber\\
&&\quad{} + 290.526 \biggl(\frac{\alpha_s}{\pi}\biggr)^{\!\!3}
- 474.036 \biggl(\frac{\alpha_s}{\pi}\biggr)^{\!\!4}
+ \cdots\biggr]\,,
\nonumber\\
&&C_m(\tau,\mu_{\tau0}) = \frac{3}{4 \pi^2 \tau^2} \biggl[1
+ 8.38649 \frac{\alpha_s}{\pi}
+ 0.125563 \biggl(\frac{\alpha_s}{\pi}\biggr)^{\!\!2}
\nonumber\\
&&\quad{} + 316.721 \biggl(\frac{\alpha_s}{\pi}\biggr)^{\!\!3}
- 307.680 \biggl(\frac{\alpha_s}{\pi}\biggr)^{\!\!4}
+ \cdots\biggr]\,,
\nonumber\\
&&C_{m^2}(\tau,\mu_{\tau0}) = - \frac{3}{8 \pi^2 \tau} \biggl[1
+ 2 \frac{\alpha_s}{\pi}
+ 19.8785 \biggl(\frac{\alpha_s}{\pi}\biggr)^{\!\!2}
\nonumber\\
&&\quad{} + 138.384 \biggl(\frac{\alpha_s}{\pi}\biggr)^{\!\!3}
+ 1579.23 \biggl(\frac{\alpha_s}{\pi}\biggr)^{\!\!4}
+ \cdots\biggr]\,.
\label{Ctnna}
\end{eqnarray}
Comparing these results with~(\ref{Ctnum}) we see that naive nonabelianization does not work well here.
The reason for this is unknown;
but the reason why naive nonabelianization works usually rather well is also unknown.

The functions $S_n(u)$ have poles at $u>0$, so that the integrals~(\ref{hatA}) are ill-defined.
Let's say there is a renormalon pole at $u_0 > 0$:
\begin{equation}
S_n(u) \sim \frac{r}{u_0 - u}\,.
\label{pole}
\end{equation}
Then the renormalon ambiguity of $\hat{A}_n(\tau)$~(\ref{hatA}), and hence of $A_n(\tau,\mu)$~(\ref{RG1}),
can be estimated by the residue of the integrand:
\begin{equation}
\Delta A_n(\tau,\mu) = \frac{C_F}{\beta_0} r \biggl(\frac{\Lambda_{\overline{\text{MS}}}}{\mu_\tau}\biggr)^{\!\!2 u_0}
= \frac{C_F}{\beta_0} r \biggl(e^{\gamma_E+5/6} \Lambda_{\overline{\text{MS}}} \frac{\tau}{2}\biggr)^{\!\!2 u_0}\,.
\label{Delta}
\end{equation}
For example, one can choose the principal-value prescription,
i.e.\ cut out the interval $[u_0-\delta,u_0+\delta]$, $\delta\to0$.
But if one changes the prescription, e.g., cuts the interval $[u_0-\delta,u_0+2\delta]$,
then the value of the integral changes by a quantity of the order of the residue.
The renormalon contribution~(\ref{pole}) expanded in $u$ and substituted into~(\ref{hatA})
produces the contribution to the perturbative series
\begin{equation}
\hat{A}_n(\tau) = 1 + \frac{C_F}{\beta_0} \sum_{l=1}^\infty a_l b^l + \mathcal{O}\biggl(\frac{1}{\beta_0^2}\biggr)\,,\quad
a_l \sim r \frac{(l-1)!}{u_0^l}
\label{asym}
\end{equation}
with the coefficients growing factorially.
If $b \ll 1$ then the series terms first decrease, then reach the minimum and start to grow.
It seems reasonable to sum such an asymptotic series up to the minimum term,
and to declare this term to be the ambiguity of the result.
The series terms $a_l b^l \sim r (l b/(e u_0))^l$ reach the minimum at $l \sim u_0/b$;
it is equal to~(\ref{Delta}).

The functions $S_n(u)$ ($n \in [0,2]$) have UV renormalon poles at $u=\frac{1}{2}$:
\begin{equation}
S_n(u) \sim e^{-\gamma_E} \frac{4}{\frac{1}{2} - u}\,.
\label{UVren}
\end{equation}
This gives the renormalon ambiguities of $A_n(\tau,\mu)$, and hence of the correlator $\Pi(\tau,\mu)$:
\begin{equation}
\frac{\Delta\Pi(\tau,\mu)}{\Pi(\tau,\mu)} = 2 \frac{C_F}{\beta_0} e^{5/6} \Lambda_{\overline{\text{MS}}} \tau
= - \Delta\bar{\Lambda}\,\tau\,,\quad
\Delta\bar{\Lambda} = - 2 \frac{C_F}{\beta_0} e^{5/6} \Lambda_{\overline{\text{MS}}}
\label{UVren0}
\end{equation}
is the UV renormalon ambiguity of $\bar{\Lambda}$,
the energy of the ground-state meson ($B$, $B^*$) in HQET~\cite{Beneke:1994sw}.
HQET energies are measured with respect to the heavy quark pole mass $M$.
This zero level is fuzzy due to the infrared (IR) renormalon $u=\frac{1}{2}$ in $M$:
$\Delta\bar{\Lambda} = - \Delta M$.
One can choose some perfectly definite zero energy level,
but then the HQET Lagrangian will contain the residual mass term~\cite{Falk:1992fm},
and it will have an IR renormalon ambiguity.
The correlator can be written as
\begin{equation}
\Pi(\tau,\mu) = \sum_i c_i e^{- \bar{\Lambda}_i \tau},
\label{disp}
\end{equation}
where $\bar{\Lambda}_i$ are the energies of all intermediate states in the correlator;
all $\Delta\bar{\Lambda}_i = \Delta\bar{\Lambda}$, and this leads to~(\ref{UVren0}).
I.e. the UV renormalon ambiguity of the correlator~(\ref{UVren0})
is compensated by the IR renormalon ambiguity of the pole mass  $M$~\cite{Beneke:1994sw}.
If one changes the summing prescription for the coefficient functions,
one has to change $M$ accordingly;
the zero energy level will shift, thus shifting all $\bar{\Lambda}_i$.

IR renormalon poles of $S_0(u)$ are situated at integer $u\ge3$:
\begin{eqnarray}
&&S_0(u) \sim - e^{-6 \gamma_E} \biggl(\frac{1}{36} \frac{1}{(3-u)^2} + \frac{2}{27} \frac{1}{3-u}\biggl)\,,
\nonumber\\
&&\frac{\Delta\Pi(\tau,\mu)}{\Pi(\tau,\mu)} = - \frac{1}{864} \frac{C_F}{\beta_0} \bigl(e^{5/6} \Lambda_{\overline{\text{MS}}} \tau\bigr)^6\,.
\label{IR0}
\end{eqnarray}
This ambiguity is compensated by the UV renormalon ambiguity of vacuum condensates.
The condensates $\langle\bar{q}q\rangle$ and $\langle\bar{q}G\sigma q\rangle$ have an opposite chirality,
their contributions are $\propto P$.
The gluon condensate $\langle G^2\rangle$ does not contribute to the correlator at 1 loop.
It is easy to understand it in coordinate space in the fixed-point gauge:
the HQET propagator does not interact with gluons;
the $G^2$ correction to the massless-quark propagator $S(x,0)$ vanishes when averaged over the vacuum~\cite{Shifman:1982zt}.
The first condensates which can contribute are the $D=6$ condensates $\langle(\bar{q}q)^2\rangle$,
therefore the first IR renormalon is at $u=3$~\cite{Beneke:1994sw}.
The functions $S_{1,2}(u)$ have IR renormalon poles at integer $u\ge1$:
\begin{eqnarray}
S_1(u) \sim e^{-2 \gamma_E} \frac{6}{1-u}
- e^{-4 \gamma_E} \biggl(\frac{3}{2} \frac{1}{(2-u)^2} + \frac{9}{2} \frac{1}{2-u}\biggr) + \cdots,
\nonumber\\
S_2(u) \sim e^{-2 \gamma_E} \frac{3}{1-u}
- e^{-4 \gamma_E} \biggl(\frac{1}{2} \frac{1}{(2-u)^2} + \frac{9}{4} \frac{1}{2-u}\biggr) + \cdots.
\label{IRS12}
\end{eqnarray}

An explicit example of such cancellation was discussed, e.g., in~\cite{Grozin:2004ez},
where the correlator of heavy-heavy quark currents at small $q^2$ at the order $1/\beta_0$ has been considered.
The coefficient function of the unit operator contains an IR renormalon at $u=2$.
The gluon condensate can be represented as the sum of the purely soft contribution
(the gluon momentum $<\lambda$) which is not calculable perturbatively
and the hard tail (the momentum $>\lambda$) which can be calculated at the order $1/\beta_0$.
The last contribution contains an UV renormalon pole at $u=2$;
its residue multiplied by the coefficient function of the gluon condensate
cancels the IR renormalon pole of the coefficient function of the unit operator.
Such an analysis of UV renormalons of dimension 6 4-quark condensates has not been done,
therefore, we cannot check the cancellation of residues at $u=3$ in our problem explicitly.

The bare coefficient function $C^0_{\sum m_i^2}$ in the large $\beta_0$ limit can be written
in the form similar to~(\ref{Ab}):
\begin{widetext}
\begin{equation}
C_{\sum m_i^2,0}(\tau) = - \frac{16}{3} \frac{N_c C_F T_F}{(4\pi)^{d/2}} \biggl(\frac{2}{\tau}\biggr)^{\!\!1-2\varepsilon} \biggl\{
\frac{1}{\beta_0^2} \sum_{l=1}^\infty \frac{F_\Sigma(\varepsilon,l\varepsilon)}{l}
\biggl[\frac{\beta_0 g_0^2}{(4\pi)^{d/2}} \biggl(\frac{\tau e^{\gamma_E}}{2}\biggr)^{\!\!2\varepsilon}
e^{-\gamma_E\varepsilon} \frac{D(\varepsilon)}{\varepsilon}\biggr]^l
+ \mathcal{O}\biggl(\frac{1}{\beta_0^3}\biggr)\biggr\}\,,
\label{bare4}
\end{equation}
where
\begin{equation}
F_\Sigma(\varepsilon,u) =
\frac{(3-2\varepsilon)^2 u (\varepsilon-u) \Gamma(2-2\varepsilon) e^{-\gamma_E(\varepsilon+2u)}}{3 (1+u) \Gamma(2-\varepsilon) \Gamma(1+\varepsilon)}
\biggl[\frac{(2-\varepsilon) \Gamma(1-u)}{(1+u) (1-\varepsilon-u) \Gamma(2-\varepsilon+u)}
- \frac{(\varepsilon+u) I(1-\varepsilon+u,\varepsilon)}{\Gamma(1-\varepsilon) \Gamma(1+2\varepsilon+2u)}\biggr]\,.
\label{FS}
\end{equation}
The function $F_\Sigma(\varepsilon,u)$ has the properties $F_\Sigma(\varepsilon,\varepsilon) = F_\Sigma(\varepsilon,0) = 0$.
The former one guarantees that there is no $l=1$ term in the sum~(\ref{bare4}) --- such a diagram does not exist.
The later one implies that there are no $\varepsilon^{-n}$, $n\ge1$ terms in the sum.
I.e., renormalization of this coefficient function is not needed ---
there are no terms which, after multiplying be a renormalization constant, would produce an element of our series.
The renormalized result is equal to the unrenormalized one and does not depend on $\mu$:
\begin{equation}
C_{\sum m_i^2}(\tau) = - \frac{16}{3} \frac{N_c C_F T_F}{(4\pi)^{d/2}} \biggl(\frac{2}{\tau}\biggr)^{\!\!1-2\varepsilon} \biggl[
\frac{1}{\beta_0^2} \sum_{l=1}^\infty \frac{F_\Sigma(\varepsilon,l\varepsilon)}{l}
\biggl(\frac{b}{\varepsilon+b}\biggr)^{\!\!l}
+ \mathcal{O}\biggl(\frac{1}{\beta_0^3}\biggr)
\biggr]\,.
\label{ren4}
\end{equation}
The result at $\varepsilon=0$ is
\begin{eqnarray}
&&C_{\sum m_i^2}(\tau) = - \frac{2 N_c C_F T_F}{3 \pi^2 \tau} \biggl[
\frac{1}{\beta_0^2} \int_0^\infty du\,e^{-u/b} S_\Sigma(u)
+ \mathcal{O}\biggl(\frac{1}{\beta_0^3}\biggr)
\biggr]\,,
\nonumber\\
&&S_\Sigma(u) = \frac{F_\Sigma(0,u)}{u} = - \frac{3 u e^{-2 \gamma_E u}}{1+u}
\biggl[\frac{2 \Gamma(1-u)}{(1+u)^2 (1-u) \Gamma(1+u)}
- \frac{u I(1+u,0)}{\Gamma(1+2u)}\biggr]\,.
\label{ren5}
\end{eqnarray}
Substituting $F(u)$~(\ref{HypExp}), we obtain
\begin{eqnarray}
&&C_{\sum m_i^2}(\tau) = - \frac{2 N_c C_F T_F}{3 \pi^2 \tau} \frac{b^2}{\beta_0^2}
\biggl[\pi^2 - 6
- 2 (18 \zeta_3 + \pi^2 - 12) b
+ 2 \biggl(54 \zeta_3 + \frac{8}{5} \pi^4 + 3 \pi^2 - 72\biggr) b^2
\nonumber\\
&&{} - 8 \biggl(360 \zeta_5 - 8 \pi^2 \zeta_3 + 66 \zeta_3 + \frac{8}{5} \pi^4 + 3 \pi^2 - 108\biggr) b^3
\nonumber\\
&&{} + 4 \biggl(3600 \zeta_5 - 900 \zeta_3^2 - 80 \pi^2 \zeta_3 + 780 \zeta_3
+ \frac{451}{63} \pi^6 + 16 \pi^4 + 30 \pi^2 - 1620\biggr) b^4
\nonumber\\
&&{} - 8 \biggl(51975 \zeta_7 - 576 \pi^2 \zeta_5 + 11016 \zeta_5 - 2700 \zeta_3^2 - 128 \pi^4 \zeta_3 - 240 \pi^2 \zeta_3 + 3060 \zeta_3
+ \frac{451}{21} \pi^6 + 48 \pi^4 + 90 \pi^2 - 6480\biggr) b^5
\nonumber\\
&&{} - 8 \biggl(16632 \zeta_{5,3} - 363825 \zeta_7 + 206136 \zeta_3 \zeta_5 + 4032 \pi^2 \zeta_5 - 78624 \zeta_5
- 2240 \pi^2 \zeta_3^2 + 19740 \zeta_3^2 + 896 \pi^4 \zeta_3 + 1680 \pi^2 \zeta_3 - 26460 \zeta_3
\nonumber\\
&&\quad{} - \frac{5269}{75} \pi^8 - \frac{451}{3} \pi^6 - 336 \pi^4 - 630 \pi^2 + 60480\biggr) b^6
+ \cdots\biggr]
+ \mathcal{O}\biggl(\frac{1}{\beta_0^3}\biggr)\,.
\label{CS}
\end{eqnarray}
\end{widetext}
Re-expressing $b=b(\mu_\tau)$ via $b_0 = b(\mu_{\tau0})$ ($b^{-1} = b_0^{-1} - \frac{5}{3}$),
we reproduce the 4-loop $N_c C_F T_F^2 n_f$ term~(\ref{sm2}).
Numerically at $\mu=\mu_{\tau0}$
\begin{eqnarray}
&&C_{\sum m_i^2}(\tau) = - \frac{\pi^2 - 6}{12 \pi^2 \tau} \biggl(\frac{\alpha_s}{\pi}\biggr)^{\!\!2}
\biggl[1 - 14.0597 \frac{\alpha_s}{\pi}
\nonumber\\
&&{} + 217.518 \biggl(\frac{\alpha_s}{\pi}\biggr)^{\!\!2} + \cdots\biggr]\,.
\label{C3tnna}
\end{eqnarray}
Comparing with~(\ref{C3tnum}) we see that naive nonabelianization works reasonably well in this case.
The function $S_\Sigma(u)$ has no UV renormalon pole at $u=\frac{1}{2}$;
IR renormalon poles are situated at integer $u\ge2$:
\begin{equation}
S_\Sigma(u) \sim \frac{1}{12} \frac{1}{(2-u)^2} + \frac{13}{72} \frac{1}{2-u}\,.
\label{IRSSigma}
\end{equation}

\subsection{Results for the spectral density}

Similarly to~(\ref{A0}) we have
\begin{eqnarray}
&&\tilde{A}_{n0}(\omega) = \frac{R_{m^n,0}(\omega)}{R_{m^n,0}^{(1)}(\omega)}
\nonumber\\
&&{} = 1 + \frac{C_F}{\beta_0} \sum_{l=1}^\infty \frac{\tilde{F}_n(\varepsilon,l\varepsilon)}{l}
\biggl[\frac{b}{\varepsilon+b} \biggl(\frac{\mu}{2\omega}\biggr)^{\!\!2\varepsilon} D(\varepsilon)\biggr]^l
\nonumber\\
&&{} + \mathcal{O}\biggl(\frac{1}{\beta_0^2}\biggr)
\label{Ar0}
\end{eqnarray}
(where the 1-loop term $R_{m^n,0}^{(1)}(\omega)$ is finite at $\varepsilon\to0$),
\begin{equation}
\tilde{F}_n(\varepsilon,u) = \frac{\Gamma(3-n-2\varepsilon)}{\Gamma(3-n-2u-2\varepsilon)} e^{2 \gamma_E u} F_n(\varepsilon,u)\,.
\label{Ft}
\end{equation}
It is convenient to set $\mu = \mu_\omega$:
\begin{equation}
\mu_\omega = 2\omega D(\varepsilon)^{-1/(2\varepsilon)} \to 2\omega e^{-5/6}\,,
\label{muw}
\end{equation}
then
\begin{equation}
\tilde{A}_n(\omega) = 1 + \frac{C_F}{\beta_0} \sum_{l=1}^\infty \frac{\tilde{F}_n(\varepsilon,l\varepsilon)}{l}
\biggl(\frac{b}{\varepsilon+b}\biggr)^{\!\!l}
+ \mathcal{O}\biggl(\frac{1}{\beta_0^2}\biggr)\,.
\label{Amuw}
\end{equation}
Naturally, $\tilde{F}_n(-b,0) = F_n(-b,0)$ gives~(\ref{gamma}) the same anomalous dimension $\gamma_n$~(\ref{gamman});
similarly to~(\ref{RG1}) and~(\ref{hatA}),
\begin{eqnarray}
&&\tilde{A}_n(\omega,\mu) = \hat{\tilde{A}}_n(\omega) \biggl(\frac{\alpha_s(\mu)}{\alpha_s(\mu_\omega)}\biggr)^{\!\!\gamma_{n,0}/(2\beta_0)}
K_n(\alpha_s(\mu))\,,
\label{RG5}\\
&&\hat{\tilde{A}}_n(\omega) =
1 + \frac{C_F}{\beta_0} \int_0^\infty du\,e^{-u/b} \tilde{S}_n(u) + \mathcal{O}\biggl(\frac{1}{\beta_0^2}\biggr)\,,
\label{hatAw}\\
&&\tilde{S}_n(u) = \frac{\tilde{F}_n(0,u)-\tilde{F}_n(0,0)}{u}
\label{Stil}
\end{eqnarray}
(here $b = b(\mu_\omega)$).
We obtain $\hat{\tilde{A}}_n(\omega)$ (similar to (\ref{Cex})):
\begin{widetext}
\begin{eqnarray}
&&\hat{\tilde{A}}_0(\omega) - \hat{A}_0(\tau) = C_F \frac{b}{\beta_0} \biggl[9
+ 3 \biggl(\pi^2 + \frac{25}{2}\biggr) b
- \biggl(160 \zeta_3 + \frac{8}{9} \pi^4 + 7 \pi^2 - \frac{451}{2}\biggr) b^2
\nonumber\\
&&\quad{} + \biggl(\frac{80}{3} \pi^2 \zeta_3 - 340 \zeta_3 + \frac{31}{3} \pi^4 - \frac{173}{2} \pi^2 + \frac{7137}{4}\biggr) b^3
\nonumber\\
&&\quad{} - \biggl(\frac{59904}{5} \zeta_5 - 1536 \zeta_3^2 - \frac{1232}{3} \pi^2 \zeta_3 + 3000 \zeta_3
  + \frac{16}{3} \pi^6 - \frac{154}{15} \pi^4 + 1013 \pi^2 - \frac{35169}{2}\biggr) b^4
\nonumber\\
&&\quad{} + \biggl(5376 \pi^2 \zeta_5 - 29904 \zeta_5 - 6080 \zeta_3^2 - \frac{2336}{9} \pi^4 \zeta_3 + 2120 \pi^2 \zeta_3 - 35660 \zeta_3
  + \frac{5476}{63} \pi^6 + 95 \pi^4 - \frac{26185}{2} \pi^2 + \frac{832005}{4}\biggr) b^5
\nonumber\\
&&\quad{} - \biggl(\frac{11918880}{7} \zeta_7 - 417792 \zeta_3 \zeta_5 - 55008 \pi^2 \zeta_5 + 322128 \zeta_5
  + \frac{27520}{3} \pi^2 \zeta_3^2 - 37280 \zeta_3^2 - \frac{7472}{3} \pi^4 \zeta_3 - 28440 \pi^2 \zeta_3 + 510660 \zeta_3
\nonumber\\
&&\qquad{} + \frac{69184}{945} \pi^8 - \frac{2612}{63} \pi^6 - 1237 \pi^4 + \frac{381435}{2} \pi^2 - \frac{11498895}{4}\biggr) b^6
+ \cdots \biggr]\,,
\label{HatR0}\\
&&\hat{\tilde{A}}_1(\omega) - \hat{A}_1(\tau) = C_F \frac{b}{\beta_0} \biggl[12
+ 2 \biggl(\frac{\pi^2}{3} + 19\biggr) b
- 2 \biggl(64 \zeta_3 + \frac{4}{9} \pi^4 + \frac{19}{3} \pi^2 - 97\biggr) b^2
\nonumber\\
&&\quad{} + \biggl(\frac{80}{3} \pi^2 \zeta_3 - 256 \zeta_3 + \frac{134}{15} \pi^4 - 97 \pi^2 + 1383\biggr) b^3
\nonumber\\
&&\quad{} - 2 \biggl(\frac{21504}{5} \zeta_5 - 768 \zeta_3^2 - \frac{400}{3} \pi^2 \zeta_3 + 1248 \zeta_3
  + \frac{8}{3} \pi^6 - \frac{38}{15} \pi^4 + 461 \pi^2 - 6351\biggr) b^4
\nonumber\\
&&\quad{} + \biggl(5376 \pi^2 \zeta_5 - 14016 \zeta_5 - 8320 \zeta_3^2 - \frac{2336}{9} \pi^4 \zeta_3 + 1520 \pi^2 \zeta_3 - 29120 \zeta_3
  + \frac{14072}{189} \pi^6 + \frac{194}{3} \pi^4 - 10585 \pi^2 + 142875\biggr) b^5
\nonumber\\
&&\quad{} - \biggl(\frac{8314560}{7} \zeta_7 - 417792 \zeta_3 \zeta_5 - 20928 \pi^2 \zeta_5 + 212544 \zeta_5
  + \frac{27520}{3} \pi^2 \zeta_3^2 - 26240 \zeta_3^2 - \frac{7904}{3} \pi^4 \zeta_3 - 23280 \pi^2 \zeta_3 + 397440 \zeta_3
\nonumber\\
&&\qquad{} + \frac{69184}{945} \pi^8 + \frac{760}{21} \pi^6 - 922 \pi^4 + 142875 \pi^2 - 1901745\biggr) b^6
+ \cdots \biggr]\,,
\label{HatR1}\\
&&\hat{\tilde{A}}_2(\omega) - \hat{A}_2(\tau) = C_F \frac{b}{\beta_0} \biggl[ - 3 \pi^2 b
- (48 \zeta_3 - \pi^2) b^2
+ 3 \biggl(8 \zeta_3 + \frac{\pi^4}{5} - \frac{19}{2} \pi^2\biggr) b^3
\nonumber\\
&&\quad{} - \biggl(\frac{6912}{5} \zeta_5 - 144 \pi^2 \zeta_3 + 912 \zeta_3 + \frac{2}{5} \pi^4 + 81 \pi^2\biggr) b^4
\nonumber\\
&&\quad{} + \biggl(1152 \zeta_5 + 1920 \zeta_3^2 - 120 \pi^2 \zeta_3 - 3240 \zeta_3
  - \frac{20}{7} \pi^6 + 19 \pi^4 - \frac{3045}{2} \pi^2\biggr) b^5
\nonumber\\
&&\quad{} - \biggl(\frac{829440}{7} \zeta_7 - 12960 \pi^2 \zeta_5 + 65664 \zeta_5
  + 1920 \zeta_3^2 + 144 \pi^4 \zeta_3 - 6840 \pi^2 \zeta_3 + 73080 \zeta_3
  - \frac{20}{7} \pi^6 - 81 \pi^4 + \frac{27855}{2} \pi^2\biggr) b^6
\nonumber\\
&&\quad{} + \cdots \biggr]\,.
\label{HatR2}
\end{eqnarray}
Here
\begin{equation}
\omega \tau = e^{-\gamma_E}\,,
\label{omegatau}
\end{equation}
so that $\mu_\omega = \mu_\tau$;
both $\hat{\tilde{A}}_n(\omega)$ and $\hat{A}_n(\tau)$
are series in $b = b(\mu_\omega) = b(\mu_\tau)$.

Setting $\mu=\mu_{\omega0}$~(\ref{muw0}) ($L = \frac{5}{6}$) we obtain
\begin{eqnarray}
&&\tilde{A}_0(\omega,\mu_\omega) = 1 + C_F \frac{b_0}{\beta_0} \biggl[
\frac{4}{3} \pi^2 + 17
- \biggl(24 \zeta_3 - \frac{23}{9} \pi^2 - \frac{1849}{24}\biggr) b_0
- \biggl(138 \zeta_3 - \frac{8}{15} \pi^4 + \frac{59}{27} \pi^2 - \frac{99901}{216}\biggr) b_0^2
\nonumber\\
&&{} - \biggl(960 \zeta_5 - 48 \pi^2 \zeta_3 + \frac{3267}{4} \zeta_3
  - \frac{533}{120} \pi^4 + \frac{18233}{162} \pi^2 - \frac{722267}{192}\biggr) b_0^3
\nonumber\\
&&{} - \biggl(\frac{53342}{5} \zeta_5 - 576 \zeta_3^2 - 560 \pi^2 \zeta_3 + \frac{115373}{18} \zeta_3
  - \frac{2176}{945} \pi^6 - \frac{1121}{36} \pi^4 + \frac{410639}{243} \pi^2 - \frac{489474263}{12960}\biggr) b_0^4
\nonumber\\
&&{} - \biggl(92400 \zeta_7 - 6400 \pi^2 \zeta_5 + \frac{551425}{6} \zeta_5
  - 8399 \zeta_3^2 + 32 \pi^4 \zeta_3 - \frac{16360}{3} \pi^2 \zeta_3 + \frac{15220631}{216} \zeta_3
\nonumber\\
&&\quad{} - \frac{16865}{567} \pi^6 - \frac{330821}{1296} \pi^4 + \frac{35328815}{1458} \pi^2 - \frac{42239102467}{93312}\biggr) b_0^5
\nonumber\\
&&{} - \biggl(25344 \zeta_{5,3} + \frac{10620906}{7} \zeta_7 - 103680 \zeta_3 \zeta_5 - 106720 \pi^2 \zeta_5 + \frac{15887005}{18} \zeta_5
  + 5760 \pi^2 \zeta_3^2 - \frac{687115}{7} \zeta_3^2
\nonumber\\
&&\quad{} + \frac{19599}{35} \pi^4 \zeta_3 - \frac{576760}{9} \pi^2 \zeta_3 + \frac{1502038795}{1512} \zeta_3
\nonumber\\
&&\quad{} - \frac{159904}{4725} \pi^8 - \frac{964501}{3969} \pi^6 - \frac{128965573}{45360} \pi^4 + \frac{1629822295}{4374} \pi^2
  - \frac{1374603729533}{217728}\biggr) b_0^6
+ \cdots \biggr]
\label{lbR0}\\
&&\tilde{A}_1(\omega,\mu_\omega) = 1 + C_F \frac{b_0}{\beta_0} \biggl[
4 \biggl(\frac{\pi^2}{3} + 6\biggr)
- \biggl(24 \zeta_3 - \frac{2}{9} \pi^2 - \frac{1097}{12}\biggr) b_0
- \biggl(100 \zeta_3 - \frac{8}{15} \pi^4 + \frac{422}{27} \pi^2 - \frac{48805}{108}\biggr) b_0^2
\nonumber\\
&&{} - \biggl(960 \zeta_5 - 48 \pi^2 \zeta_3 + \frac{1091}{2} \zeta_3
  - \frac{181}{60} \pi^4 + \frac{13837}{81} \pi^2 - \frac{2804531}{864}\biggr) b_0^3
\nonumber\\
&&{} - \biggl(\frac{36284}{5} \zeta_5 - 576 \zeta_3^2 - 416 \pi^2 \zeta_3 + \frac{42253}{9} \zeta_3
  - \frac{2176}{945} \pi^6 - \frac{1493}{90} \pi^4 + \frac{438986}{243} \pi^2 - \frac{64308541}{2160}\biggr) b_0^4
\nonumber\\
&&{} - \biggl(92400 \zeta_7 - 6400 \pi^2 \zeta_5 + \frac{142625}{3} \zeta_5
  - 6238 \zeta_3^2 + 32 \pi^4 \zeta_3 - \frac{10960}{3} \pi^2 \zeta_3 + \frac{5986231}{108} \zeta_3
\nonumber\\
&&\quad{} - \frac{9794}{567} \pi^6 - \frac{92485}{648} \pi^4 + \frac{15753115}{729} \pi^2 - \frac{15604351747}{46656}\biggr) b_0^5
\nonumber\\
&&{} - \biggl(25344 \zeta_{5,3} + \frac{7012212}{7} \zeta_7 - 103680 \zeta_3 \zeta_5 - 72640 \pi^2 \zeta_5 + \frac{4240925}{9} \zeta_5
  + 5760 \pi^2 \zeta_3^2 - \frac{460310}{7} \zeta_3^2
\nonumber\\
&&\quad{} + \frac{14558}{35} \pi^4 \zeta_3 - \frac{422320}{9} \pi^2 \zeta_3 + \frac{589984715}{756} \zeta_3
\nonumber\\
&&\quad{} - \frac{159904}{4725} \pi^8 - \frac{161530}{3969} \pi^6 - \frac{42604613}{22680} \pi^4 + \frac{651102575}{2187} \pi^2
  - \frac{1451860425719}{326592}\biggr) b_0^6
+ \cdots \biggr]
\label{lbR1}\\
&&\tilde{A}_2(\omega,\mu_\omega) = 1 + C_F \frac{b_0}{\beta_0} \biggl[
6
- \biggl(3 \pi^2 - \frac{229}{8}\biggr) b_0
- \biggl(30 \zeta_3 + 9 \pi^2 - \frac{5329}{72}\biggr) b_0^2
- \biggl(\frac{633}{2} \zeta_3 - \frac{21}{20} \pi^4 + 97 \pi^2 - \frac{88673}{96}\biggr) \frac{b_0^3}{2}
\nonumber\\
&&{} - \biggl(\frac{6426}{5} \zeta_5 - 144 \pi^2 \zeta_3 + \frac{2335}{2} \zeta_3
  - \frac{71}{20} \pi^4 + \frac{2789}{9} \pi^2 - \frac{15229703}{4320}\biggr) b_0^4
\nonumber\\
&&{} - \biggl(\frac{19425}{2} \zeta_5 - 2157 \zeta_3^2 - 1080 \pi^2 \zeta_3 + \frac{223037}{24} \zeta_3
  + \frac{181}{63} \pi^6 - \frac{4663}{144} \pi^4 + \frac{55055}{18} \pi^2 - \frac{1105270627}{31104}\biggr) b_0^5
\nonumber\\
&&{} - \biggl(\frac{816318}{7} \zeta_7 - 12960 \pi^2 \zeta_5 + \frac{209535}{2} \zeta_5
  - \frac{136065}{7} \zeta_3^2 + \frac{5037}{35} \pi^4 \zeta_3 - 11640 \pi^2 \zeta_3 + \frac{6166355}{56} \zeta_3
\nonumber\\
&&\quad{} + \frac{11345}{441} \pi^6 - \frac{1561591}{5040} \pi^4 + \frac{1905235}{54} \pi^2 - \frac{30732738973}{72576}\biggr) b_0^6
+ \cdots \biggr]
\label{lbR2}
\end{eqnarray}
\end{widetext}
(here $b_0 = b(\mu_{\omega0})$).
Substituting $\beta_0 \to - \frac{4}{3} T_F n_f$,
we reproduce all $C_F (T_F n_f)^{l-1} \alpha_s(\mu_{\omega0})^l$ terms in (\ref{r0}--\ref{r2}),
and add further terms up to 8 loops (more such terms can be added if desired).
Leaving $\beta_0 = \frac{11}{3} C_A - \frac{4}{3} T_F n_f$ (naive nonabelianization)
we get numerically with $n_f = 4$
\begin{eqnarray}
&&R_1(\omega,\mu_{\omega0}) = \frac{3 \omega^2}{4 \pi^2} \biggl[1
+ 10.0532 \frac{\alpha_s}{\pi}
+ 50.9824 \biggl(\frac{\alpha_s}{\pi}\biggr)^{\!\!2}
\nonumber\\
&&\quad{} + 473.098 \biggl(\frac{\alpha_s}{\pi}\biggr)^{\!\!3}
+ 5051.24 \biggl(\frac{\alpha_s}{\pi}\biggr)^{\!\!4}
+ \cdots\biggr]\,,
\nonumber\\
&&R_m(\omega,\mu_{\omega0}) = \frac{3 \omega}{4 \pi^2} \biggl[1
+ 12.3865 \frac{\alpha_s}{\pi}
+ 44.9726 \biggl(\frac{\alpha_s}{\pi}\biggr)^{\!\!2}
\nonumber\\
&&\quad{} + 331.866 \biggl(\frac{\alpha_s}{\pi}\biggr)^{\!\!3}
+ 2327.26 \biggl(\frac{\alpha_s}{\pi}\biggr)^{\!\!4}
+ \cdots\biggr]\,,
\nonumber\\
&&R_{m^2}(\omega,\mu_{\omega0}) = - \frac{3}{8 \pi^2} \biggl[1
+ 2 \frac{\alpha_s}{\pi}
- 0.683204 \biggl(\frac{\alpha_s}{\pi}\biggr)^{\!\!2}
\nonumber\\
&&\quad{} - 73.6028 \biggl(\frac{\alpha_s}{\pi}\biggr)^{\!\!3}
- 469.965 \biggl(\frac{\alpha_s}{\pi}\biggr)^{\!\!4}
+ \cdots\biggr]\,.
\label{Rwnna}
\end{eqnarray}
Comparing these results with~(\ref{Rwnum}) we see that naive nonabelianization does not work well here,
just like the case of $C_{m^n}(\tau,\mu_{\tau0})$~(\ref{Ctnna}).

The functions $\tilde{S}_n(u)$ ($n \in [0,2]$) have UV renormalon poles at $u=\frac{1}{2}$:
\begin{equation}
\tilde{S}_n(u) \sim \frac{4 (2-n)}{\frac{1}{2} - u}\,,\quad
\frac{\Delta R_{m^n}(\omega,\mu)}{R_{m^n}(\omega,\mu)} = (2-n) \frac{\Delta\bar{\Lambda}}{\omega}\,.
\label{UVrho}
\end{equation}
This is not surprising: $R_{m^n}(\omega) \propto \omega^{2-n}$, $\Delta\omega = \Delta\bar{\Lambda}$.
The first IR renormalon pole of $\tilde{S}_0(u)$ is at $u=3$:
\begin{equation}
\tilde{S}_0(u)\sim \frac{2}{3} \frac{1}{3-u}\,,\quad
\frac{\Delta\rho(\omega,\mu)}{\rho(\omega,\mu)} = \frac{1}{96} \frac{C_F}{\beta_0}
\biggl(e^{5/6} \frac{\Lambda_{\overline{\text{MS}}}}{\omega}\biggr)^{\!\!6}\,.
\label{IRrho}
\end{equation}
This ambiguity is compensated by the UV renormalon ambiguity
of the condensates $\langle(\bar{q}q)^2\rangle$ of dimension 6 in OPE.
The functions $\tilde{S}_{1,2}(u)$ have IR renormalon poles at $u = 2$, 3\ldots
(the factor in~(\ref{Ft}) eliminates simple poles in~(\ref{IRS12}) and converts double poles to simple ones):
\begin{eqnarray}
&&\tilde{S}_1(u) \sim - \frac{6}{2-u}\,,\quad
\frac{\Delta R_m(\omega,\mu)}{R_m(\omega,\mu)} = - \frac{3}{2} \frac{C_F}{\beta_0}
\biggl(e^{5/6} \frac{\Lambda_{\overline{\text{MS}}}}{\omega}\biggr)^{\!\!4}\,,
\label{IR1}\\
&&\tilde{S}_2(u) \sim \frac{6}{2-u}\,,\quad
\frac{\Delta R_{m^2}(\omega,\mu)}{R_{m^2}(\omega,\mu)} = - 3 \frac{C_F}{\beta_0}
\biggl(e^{5/6} \frac{\Lambda_{\overline{\text{MS}}}}{\omega}\biggr)^{\!\!4}\,.
\label{IR2}
\end{eqnarray}
The IR renormalon ambiguity of $m R_m(\omega,\mu)$~(\ref{IR1}) is compensated by the UV renormalon ambiguity
of the condensate $\langle\bar{q}G\sigma q\rangle$ which has the same chirality (the contribution $\propto P$).
The IR renormalon ambiguity of $m^2 R_{m^2}(\omega,\mu)$~(\ref{IR2}) is compensated by the UV renormalon ambiguity
of the gluon condensate in the $m^2 \langle G^2\rangle$ contribution,
as well as of the condensates $\langle(\bar{q}q)^2\rangle$.

For the $\sum m_i^2$ contribution we have, similarly to~(\ref{Ft}),
\begin{equation}
\tilde{F}_\Sigma(\varepsilon,u) = \frac{e^{2 \gamma_E u}}{\Gamma(1-2\varepsilon-2u)} F_\Sigma(\varepsilon,u)\,,
\label{FStilda}
\end{equation}
and, similarly to~(\ref{CS}), we obtain
\begin{widetext}
\begin{eqnarray}
&&R_{\sum m_i^2}(\omega) - \tau C_{\sum m_i^2}(\tau) = - \frac{2 N_c C_F T_F}{3 \pi^2} \frac{b^4}{\beta_0^2} \biggl[
- 2 \pi^2 (\pi^2 - 6)
+ 8 (10 \pi^2 \zeta_3 + 48 \zeta_3 + \pi^4 - 12 \pi^2) b
\nonumber\\
&&{} + 4 (1440 \zeta_3^2 - 100 \pi^2 \zeta_3 - 960 \zeta_3 - 5 \pi^6 - 12 \pi^4 + 240 \pi^2) b^2
\nonumber\\
&&{} + 8 (3024 \pi^2 \zeta_5 + 3456 \zeta_5 - 4320 \zeta_3^2 - 146 \pi^4 \zeta_3 - 60 \pi^2 \zeta_3 + 5760 \zeta_3
  + 15 \pi^6 + 42 \pi^4 - 1080 \pi^2) b^3
\nonumber\\
&&{} + 8 \biggl(274176 \zeta_3 \zeta_5 - 21168 \pi^2 \zeta_5 - 48384 \zeta_5 - 6020 \pi^2 \zeta_3^2 + 23520 \zeta_3^2
  + 1022 \pi^4 \zeta_3 + 2940 \pi^2 \zeta_3 - 60480 \zeta_3
\nonumber\\
&&\quad{} - \frac{2162}{45} \pi^8 - 95 \pi^6 - 378 \pi^4 + 11340 \pi^2\biggr) b^4 + \cdots \biggr]\,.
\label{RS}
\end{eqnarray}
\end{widetext}
Here $\omega$ and $\tau$ are related by~(\ref{omegatau}).

Re-expressing $b=b(\mu_\omega)$ via $b_0 = b(\mu_{\omega0})$, we reproduce the 4-loop $N_c C_F T_F^2 n_f$ term~(\ref{RG3}).
Numerically at $\mu=\mu_{\tau0}$
\begin{eqnarray}
&&R_{\sum m_i^2}(\omega) = - \frac{\pi^2 - 6}{12 \pi^2} \biggl(\frac{\alpha_s}{\pi}\biggr)^{\!\!2}
\biggl[1 - 14.0597 \frac{\alpha_s}{\pi}
\nonumber\\
&&{} + 217.518 \biggl(\frac{\alpha_s}{\pi}\biggr)^{\!\!2} + \cdots\biggr]\,.
\label{R3wnna}
\end{eqnarray}
Comparing with~(\ref{R3wnum}) we see that naive nonabelianization works reasonably well in this case.
The Borel image $\tilde{S}_\Sigma(u)$~(\ref{ren5}) has IR renormalon poles at positive integer $u$ starting from $u=2$.

\section{Conclusion}
\label{S:Conc}

We have calculated the perturbative contribution to the correlator of HQET heavy-light quark currents
(expanded in light-quark masses up to quadratic terms) with the 4-loop accuracy.
They can be used in QCD sum rules for heavy-light mesons to calculate $f_{B^{(*)}}$, $M_{B^{(*)}}$.
The light-quark mass corrections are useful for flavor $SU(3)$ violations
like $f_{B_s^{(*)}}/f_{B^{(*)}}$, $M_{B_s^{(*)}}-M_{B^{(*)}}$.
The results for the $P=-1$ current can be used to obtain similar quantities
for the $P$-wave $0^+$, $1^+$ mesons.
First we express the heavy-light QCD currents $\bar{Q}\Gamma q$ as the HQET currents times the matching coefficients $C_\Gamma(\mu)$
(known up to 3 loops~\cite{Bekavac:2009zc})
plus $1/M$ suppressed terms
(higher-dimensional HQET operators times corresponding matching coefficients).
These $1/M$ corrections to $f_{B^{(*)}}$, $M_{B^{(*)}}$ should be calculated separately,
maybe, also using QCD sum rules for correlators involving these higher-dimensional HQET operators.
At the leading order in $1/M$ we have 2 HQET quantities:
the matrix element $F(\mu)$ and $\bar{\Lambda}$
($\langle0|j_+(\mu)|M\rangle = F(\mu) u$, where $u$ is the Dirac wave function of the $\frac{1}{2}^+$ meson $M$
with the $0^+$ infinitely heavy antiquark, the state $|M\rangle$ is normalized non-relativistically);
$f_B = 2/\sqrt{M} C_{\gamma^0}(\mu) F(\mu)$, $f_{B^*} = 2/\sqrt{M} C_{\vec{\gamma}}(\mu) F(\mu)$,
$M_B = M_{B^*} = M+\bar{\Lambda}$.
These 2 quantities can be found from the sum rule
\begin{equation}
|F(\mu)|^2 e^{-\bar{\Lambda}\tau}
= \int_0^{\omega_c} d\omega\,\rho^{(D\le2)}(\omega,\mu) e^{-\omega\tau}
+ \Pi^{(D\ge3)}(\tau,\mu)\,,
\label{SR}
\end{equation}
where $\omega_c$ is the continuum threshold.

Of course, $\bar{\Lambda}$ is not well defined due to the renormalon at $u=\frac{1}{2}$.
In order to alleviate problems with poor convergence of the perturbative series,
it is reasonable to multiply both sides of this equation by $\exp[(M_{\text{reg}}(\mu_f)-M)\tau]$,
where $M_{\text{reg}}(\mu_f)$ is some mass having no renormalon $u=\frac{1}{2}$
but close to the on-shell mass $M$ ($M_{\text{reg}}(\mu_f)-M = \mathcal{O}(M^0)$),
such as, e.g., the potential subtracted mass $M_{\text{PS}}(\mu_f)$~\cite{Beneke:1998rk}
(there are also other mass definitions with these properties).
Then $\bar{\Lambda}_{\text{reg}}(\mu_f) = \bar{\Lambda} + M - M_{\text{reg}}(\mu_f)$ has no $u=\frac{1}{2}$ renormalon ambiguity,
and the sum rule for it will have a better behavior at large orders of perturbation theory.
Having obtained $\bar{\Lambda}_{\text{reg}}(\mu_f)$ from the sum rule at a suitable value of the factorization scale $\mu_f$,
we obtain $M_{\text{reg}}(\mu_f) = M_B - \bar{\Lambda}_{\text{reg}}(\mu_f)$,
and can obtain the $b$ quark mass in any desired scheme by comparing these mass definitions.
Quantities like $f_{B_s}/f_B$, $\bar{\Lambda}_s-\bar{\Lambda} = M_{B_s} - M_B$, etc. don't suffer from these renormalon problems.

We have also considered this correlator in the large-$\beta_0$ framework.
It is a convenient model for considering large-order behavior of perturbative series
and ambiguities of their sums.
The Borel images of the OPE coefficient functions contain UV renormalon poles at $u=\frac{1}{2}$,
related to the UV renormalon of the HQET residual energy $\bar{\Lambda}$
(or the IR renormalon of the on-shell heavy-quark mass $M$ in QCD).
They also contain IR renormalon poles at positive integer values of $u$;
the corresponding IR renormalon ambiguities are compensated by UV renormalon ambiguities
of vacuum averages of higher-dimensional operators in the OPE.
Surprisingly, naive nonabelianization~\cite{Broadhurst:1994se,Beneke:1994qe}
works very poorly for these coefficient functions.

\begin{acknowledgments}
I am grateful to R.N.~Lee for useful discussions.
The work presented in Sect.~\ref{S:4L} was supported by the Russian Science Foundation, grant number 20-12-00205.
The work presented in Sect.~\ref{S:Lb0} was supported by the Russian Ministry of Science and Higher Education.
\end{acknowledgments}

\bibliography{corr4}
\end{document}